\documentclass[11pt,english,twoside]{article}

\usepackage[T1]{fontenc}
\usepackage[latin1]{inputenc}
\usepackage[english]{babel}
\usepackage{lmodern}

\usepackage{amsmath,amssymb, cite}
\usepackage{amssymb}
\usepackage{mathrsfs}
\usepackage{amsmath,amsthm}
\usepackage[hypertex]{hyperref}
\usepackage{array,longtable}
\usepackage[dvips]{graphicx}
\usepackage[dvips]{color}
\usepackage{colortbl}
\usepackage{wrapfig}
\usepackage{bbm}

\usepackage{float}
\usepackage{graphicx}
\usepackage[dvips]{epsfig}
\usepackage{psfrag}

\newcommand{\beq}{\begin{equation}}
\newcommand{\eeq}{\end{equation}}
\newcommand{\bea}{\begin{eqnarray}}
\newcommand{\eea}{\end{eqnarray}}
\newcommand{\nn}{\nonumber}
\newcommand{\w}{\wedge}

\newcommand{\del}{\partial}

\newcommand{\bbZ}{\mathbb{Z}}
\newcommand{\bbR}{\mathbb{R}}

\DeclareMathOperator{\SU}{\mathit{SU}}

\DeclareMathOperator{\GL}{\mathit{GL}}

\DeclareMathOperator{\Spin}{\mathit{Spin}}

\newcommand{\aaa}{{\cal A}}
\newcommand{\bbb}{{\cal B}}
\newcommand{\ccc}{{\cal C}}

\newcommand{\fff}{{\cal F}}

\newcommand{\dd}{{\rm d}}
\newcommand{\ee}{{\rm e}}

\DeclareMathOperator{\re}{Re}
\DeclareMathOperator{\im}{Im}

\DeclareMathOperator{\vol}{vol}

\newcommand{\cE}{\mathcal{E}}

\newcommand{\Ggeom}{G_{\mathrm{geom}}}

\newcommand{\ov}{\overline}

\numberwithin{equation}{section}       

\begin{document}










\begin{titlepage}

\begin{center}

\rightline{\small SPhT-T09/022}

\vskip 2cm

\begin{LARGE}
   \textbf{Flux backgrounds from Twists}
\end{LARGE}

\vskip 1.2cm

\textbf{David Andriot$^{a}$, Ruben Minasian$^{b}$, Michela Petrini$^{a}$}

\vskip 0.8cm
{}$^a$ \textit{LPTHE, CNRS, UPMC Univ Paris 06\\
Bo\^ite 126, 4 Place Jussieu\\
F-75252 Paris cedex 05, France} \\
\vskip 0.4cm
{}$^{b}$\textit{Institut de Physique Th\'eorique, CEA/Saclay \\
91191 Gif-sur-Yvette Cedex, France} \\

\end{center}

\vskip 1cm

\begin{center} {\bf Abstract } \end{center}
It is well known that a constant $O(n,n,\mathbb{Z})$ transformation can relate different 
string backgrounds with $n$ commuting isometries that have very different geometric 
and topological properties. 
Here we construct discrete families of (flux) backgrounds on internal manifolds of 
different topologies by performing certain coordinate dependent $O(d,d)$ 
transformations, where $d$ is the dimension of the internal  manifold.
Our two principal examples include 
respectively the family of type IIB compactifications with D5 branes and O5 planes 
on six-dimensional nilmanifolds, and the heterotic torsional backgrounds.

\vfill
\today

\end{titlepage}


\tableofcontents


\section{Introduction}

The study of supersymmetric string backgrounds reveals fascinating connections between physical 
dualities and  geometric transitions. Quite often new duality relations have interesting geometric 
implications. It has been known for a long time that in presence of isometries there
exist discrete families of equivalent quantum field theories which can be formulated on different backgrounds. An action of constant $O(n,n,\mathbb{Z})$ matrices leaves invariant the system of the
sigma model equations of motion and Bianchi identities, while at the level of the target space
it yields very different compactification manifolds (see \cite{GPR} for a review).
This duality, which is a generalization of T-duality, is part of a larger web of string dualities 
and has a special significance due to its perturbative nature.  

Recent studies of string compactifications have produced examples of pairs of backgrounds which,
on one side, display clear geometric relations, while, on the other,  
do not seem to be connected by any simple direct duality. We have two examples in mind which will 
play an important role in the discussion of this paper. 
One example concerns type II compactifications on six-tori with fluxes. In type IIB
such configurations involve O3 planes, and can be supersymmetric provided the fluxes 
are of a suitable type \cite{GP}. Moreover the fluxes (as well as the warping) can be arranged in 
such a way as to respect some of the isometries of the torus. A sequence of two dualities 
should still leave us within IIB theory, and the outcome is a model with O5/D5 sources 
on a parallelizable nilmanifold with the first Betti number being equal to 5 or 4
\cite{KSTT, Sc}. There is a single solution involving a  nilmanifold as 
internal space  that does not seem to be  accessible by any duality. It  corresponds to a type IIB solution 
on a manifold which can be seen as an iteration of circle fibrations and has first Betti 
number $b_1(M)=3$ \cite{GMPT06}. 

The second example is the K\"ahler/non-K\"ahler transition in the heterotic compactifications with non-trivial $H$-flux \cite{S, Hu}. The two respective compactification manifolds are both locally given by a product of $K3$ surface and a two torus. While it is long believed that such transitions should exist and can connect different Calabi-Yau manifolds, the relation is established via a complicated and indirect chain of dualities involving a lift to M-theory (see \cite{DRS, BD, GoldP, FuY, BBFTY, Yi1, Yi2, BTY, Se}).\\

These examples provide us with a motivation to look for a direct transformation to relate these seemingly isolated solutions.
Since locally the compactification manifolds are of the form $\bbb \times \mathbb{T}^n$, a very naive idea is to try to make a global action depend on the coordinates on $\bbb$ in order to bring-in a connection. If this connection is non-trivial, it will be responsible for the topology change via the transformation, and the tadpole cancellation will fix the quantization condition for its curvature.

We find a special form of such a transformation that does indeed provide a direct and simple connection 
when applied to the above mentioned examples. The tools for finding the transformation are provided by the Generalized Complex Geometry \cite{H,G}. The latter has already played an important role in the string compactifications since having a Generalized Calabi-Yau structure is a necessary condition for preserving supersymmetry \cite{GMPT04,GMPT05}. Moreover, the so-called generalized tangent bundle \cite{H2}, namely the extension of the tangent bundle by the cotangent one which combines
metric and $B$-field, conveniently encodes all the global data needed to understand the action of the constant 
$O(n,n,\mathbb{Z})$ transformations. The latter are part of a more general $O(d,d)$ action\footnote{Throughout the paper, we will use $n$ for the number of isometries, and hence $O(n,n,\mathbb{Z})$ for the T-duality group, while $d$ will be denoting the dimension of the internal manifold (mostly $d=6$).} which leaves the metric on the generalized tangent bundle invariant. The transformation we design here is a special case of an $O(d,d)$ transformation which has been made coordinate dependent. Generically, the transformation is a combination of a $B$-transform, a (scaling type) transformation of (parts of) the metric and accordingly a shift in 
dilaton, a (pair of) $U(1)$ rotation(s) and a change in the connection. Since the latter provides the most easily 
distinguishable feature of related manifolds, we shall use ``twist transformation'' for shorthand. \\

We apply the ``twist transformation'' to construct families of backgrounds for the examples we mentioned above. More
precisely, we will construct a  family of O5/D5 configurations on twisted tori (iterations of torus fibrations over tori)
from a $\mathbb{T}^6$ compactification with O3 planes, and heterotic torsional backgrounds starting from a 
$K3\times \mathbb{T}^2$ compactification. These are special cases of (K\"ahler/non-K\"ahler) transitions between manifolds of vanishing Chern class. We are considering here smooth fibrations.
It would be of some interest to extend this construction for degenerating fibres and understand 
if it can give the general transitions between manifolds with a trivial canonical bundle. 

As it is clear from these examples, the claim is that the twist transformation can relate 
flux backgrounds to compactifications on Ricci flat backgrounds. This means in particular 
that in type II backgrounds it does mix NSNS and RR fields. One way of getting a mixing of these 
two sectors is by U-duality. The corresponding extensions of the Generalized Complex Geometry 
is provided by the so-called Exceptional Generalized Geometry, which
have been argued by considering the sum of higher powers of the tangent and  the cotangent bundles needed to enforce the twisting with respect to RR fields \cite{Hu2, PW, GLSW}. Our transformation on the contrary only uses the generalized tangent bundle.\\

This paper is organised as follows. In Section \ref{odd}, we define our $O(6,6)$ twist transformation, and detail its 
action both on the generalized vielbein and on the pure spinors.  Then, we consider the action of the twist 
specifically for the case of  $\mathbb{T}^2$ fibrations. 
In Section \ref{tpII}, we apply our transformation to type II compactifications. We first discuss the general 
constraints that the transformation needs to satisfy in order for it to generate new solutions. Then, we use the twist
transformation to map $\mathbb{T}^6$ backgrounds with O3 planes to a family of solutions on twisted tori 
with O5/D5 sources. In Section \ref{het}, we turn to heterotic string. After reviewing the conditions for having $\mathcal{N}=1$ solutions, we construct the pure spinors equations reproducing these conditions. Then, we show how to obtain the 
K\"ahler/non-K\"ahler transition via our transformation on the pure spinors. The 
corresponding transformation on the generalized vielbein might need an extension of the generalized tangent bundle to 
include the gauge bundle, a possibility further explored in Appendix \ref{Fhet}. Finally, 
in Section \ref{Courant}, we explore the possibility for our transformation being an automorphism 
of the Courant bracket. 

\section{Twists from $O(d,d)$ transformations}
\label{odd}

In compactifications on manifolds with a $n$-dimensional torus action, $O(n,n)$ transformations on the fibre have been successfully used
to generate new solutions of string theory/supergravity. The classical example is T-duality, which can give
rise to new geometries or to non-geometric backgrounds, depending on the form of
the $B$-field of the original solution.  A more recent example of $O(n,n)$ transformation is
the so-called $\beta$-transform:  a combination of T-duality, rotation and another T-duality. Its most popular application is  to the $AdS_5 \times S^5$ background, where it produces the supergravity dual
of $\beta$-deformed ${\cal N}=4$ Super Yang-Mills \cite{LM, MPZ, HT}. 
All these transformations act purely on the fibres of the internal manifold $M_d$.
The  purpose of this paper is to study the action of a different type of  transformation that mixes fibre and base directions.\\

$O(d,d)$ transformations appear naturally in Generalized (Complex) Geometry\footnote{In Appendix A, we
give a brief summary of the Generalized Complex Geometry we will need in this paper.} as the stabilizer of the metric on the generalized tangent bundle, $E$. Given a $d$-dimensional manifold $M$, the generalized tangent bundle $E$ is a non-trivial fibration
of $T^*M$ over $TM$
\begin{equation}
\label{eq:Edef}
   0 \longrightarrow T^*M \longrightarrow E \longrightarrow TM \longrightarrow 0 .
\end{equation}
The sections of $E$ are called generalized vectors and they can be
written locally as
\beq
\label{eq:sections}
X= v+\xi = \begin{pmatrix} v \\ \xi
\end{pmatrix} \, ,
\eeq
where $v\in TM$ and $\xi\in T^*M$. The transition functions patching the
generalized vectors between two coordinate patches $U_\alpha$ and
$U_\beta$  are
\beq
\label{eq:S-patch}
\begin{pmatrix}
v \\ \xi
\end{pmatrix}_{(\alpha)} = \begin{pmatrix} a & 0 \\ \omega a & a^{-T} \end{pmatrix}_{(\alpha\beta)}
\begin{pmatrix}
v \\ \xi
\end{pmatrix}_{(\beta)} \, .
\eeq
$a_{(\alpha\beta)}$ is an element of $\GL(d,\bbR)$, and gives the
usual patching of vectors and one-forms. To simplify notations
we set  $a^{-T}=(a^{-1})^T$.
The additional shift of the one-form gives the non trivial fibration
of $T^*M$ over $TM$. $\omega_{(\alpha\beta)}$ is a two-form such that
$\omega_{(\alpha\beta)}=-\dd\Lambda_{(\alpha\beta)}$, and it defines a  ``connective
structure'' of a gerbe.

$E$ is equipped with a natural metric, defined by the coupling of vectors
and one-forms
\beq
\label{eq:eta}
\eta(X,X) = i_v\xi  \qquad \Leftrightarrow \qquad X^T \eta X  = \frac{1}{2} \begin{pmatrix} v & \xi
\end{pmatrix} \, \begin{pmatrix} 0 & 1 \\ 1 & 0 \end{pmatrix} \,
 \begin{pmatrix} v \\ \xi
\end{pmatrix} \, .
\eeq
The metric $\eta$ is invariant under $O(d,d)$ transformations, which
act on the generalized vectors in 
the fundamental representation 
\beq
\label{eq:Oddfun}
X'=  O X = \begin{pmatrix} a & b \\ c & d \end{pmatrix}
         \begin{pmatrix} v \\ \xi \end{pmatrix} .
\eeq

However, from the patching condition \eqref{eq:S-patch}, it follows that
the structure group of $E$ is reduced to the subgroup of $O(d,d)$ given
by the semi-direct product $\Ggeom=G_B\rtimes\GL(d)$
\begin{equation}
\label{eq:ggeom}
   P = \ee^B \begin{pmatrix} a & 0 \\ 0 & a^{-T} \end{pmatrix}
       =  \begin{pmatrix} a & 0 \\ B a & a^{-T} \end{pmatrix} \, .
\end{equation}
$\GL(d)$ acts in the
usual way on the fibres of $TM$ and $T^*M$
\beq
\label{eq:diffeo}
   X \mapsto X' =
      \begin{pmatrix} a & 0 \\ 0 & a^{-T} \end{pmatrix}
      \begin{pmatrix} v \\ \xi \end{pmatrix} \, .
\eeq
The factor $G_B$ is called $B$-transform and it is
generated by the action of a two-form $B$
\beq
\label{eq:Btransform}
X \mapsto X' =  \ee^B  X =
      \begin{pmatrix} \mathbb{I} & 0 \\ B & \mathbb{I} \end{pmatrix}
\begin{pmatrix} v \\ \xi \end{pmatrix} =
\begin{pmatrix} v \\ \xi-i_v B \end{pmatrix} \, .
\eeq
The embedding
of $\Ggeom\subset O(d,d)$ is fixed by the projection $\pi:E\to TM$. It is
the subgroup which leaves the image of the related embedding $T^*M\to
E$ invariant. \\


In this  paper we shall show how we can use $\Ggeom$ to relate different string backgrounds. 
We will be mostly concerned with the case where $M$ is a torus fibration, 
$ \mathbb{T}^n\hookrightarrow M \stackrel{\pi} \longrightarrow {\bbb}$, and  consider an element of $\Ggeom=G_B\rtimes\GL(d)$  of the type
\beq
\label{eq:O(6,6)tr}
O = \begin{pmatrix} A & 0 \\ C & D \end{pmatrix}
=  \begin{pmatrix}  A_{\bbb}      & 0   & 0   & 0 \\
                    A_{\ccc}      & A_{\fff} & 0   & 0 \\
                    C_{\bbb}      & C_{\ccc} & D_{\bbb} & D_{\ccc} \\
                    C_{\ccc '} & C_{\fff} & 0   & D_{\fff} \end{pmatrix} \, .
\eeq
In the second matrix, we split the base ($\bbb$), fibre ($\fff$) and mixed elements. The $O(6,6)$ constraints reduce in this case to
\beq
A^T C+C^T A=0 \qquad \qquad  A^T D = \mathbb{I}  \ .
\eeq
This fixes the matrix $D$ to be the inverse of $A$
\beq
D=(A^T)^{-1}= \begin{pmatrix} A^{-T}_{\bbb} & - A^{-T}_{\bbb}  A_{\ccc}^T A^{-T}_{\fff}  \\ 0 & A^{-T}_{\fff}  \end{pmatrix} \, ,
\eeq
and allows to parametrise $C$ in terms of three unconstrained matrices $\tilde{C}_{\bbb}$, $\tilde{C}_{\fff}$,
$\tilde{C}_{\ccc}$,
\beq
C= \begin{pmatrix} A^{-T}_{\bbb} (\tilde{C}_{\bbb} -A^T_{\ccc} A^{-T}_{\fff} \tilde{C}_{\ccc}) &
- A^{-T}_{\bbb}(\tilde{C}^T_{\ccc}  + A^T_{\ccc} A^{-T}_{\fff} \tilde{C}_{\fff}) \\
A^{-T}_{\fff} \tilde{C}_{\ccc} & A^{-T}_{\fff} \tilde{C}_{\fff} \end{pmatrix} \, ,
\eeq
with $\tilde{C}_{\bbb}$ and $\tilde{C}_{\fff}$ anti-symmetric.

This transformation naturally combines  the fibration structure of the internal manifold with the standard symmetries of the generalized tangent bundle.   A $\mathbb{T}^n$ invariant section of $TM$ ($T^*M)$ can be considered as an element of $T\bbb \oplus \mathfrak{t}$ 
($T^*\bbb \oplus \mathfrak{t}^*$), where  $\mathfrak{t}:=\mathrm{Lie}\,\mathbb{T}^n\cong \mathbb{R}^n$.
We can now think of the generalized tangent bundle $E$ as a bundle over $\bbb \times \mathbb{T}^n$
, and interpret our transformation (\ref{eq:O(6,6)tr}) as a generalized $B$-transform. In a more conventional language this would be a combination of an ordinary $B$-transform \eqref{eq:Btransform} and a twisting of $\mathbb{T}^n$ over $\bbb$.  

We shall study when and how the transformation (\ref{eq:O(6,6)tr}) maps one string background to another. In general, two internal manifolds connected in this way  will have different topologies. Typically such topology changes are associated with large transformations, while (\ref{eq:O(6,6)tr})  is connected to the identity. The topological properties of related backgrounds are determined by the global properties of the matrices $C$ and $A_{\ccc}$.

\subsection{Action on the generalized vielbeins}\label{subgenviel}

One reason to introduce the transformation \eqref{eq:O(6,6)tr} is to map spaces that are
direct products of two manifolds, for instance $K3 \times \mathbb{T}^2$, into spaces
that are non-trivial fibrations.
To see how this is achieved we can look at the $O(d,d)$ transformations of the generalized vielbeins.\\

In Generalized Geometry the metric $g$ and the $B$-field
combine into a single object, the generalized metric
\beq
\label{eq:genmetric}
{\cal H} = \begin{pmatrix}
         g - B g^{-1} B & B g^{-1} \\
         - g^{-1} B & g^{-1}
      \end{pmatrix} \, ,
\eeq
and, as in conventional geometry, it is possible to write it in terms of generalized vielbein
\beq
\label{eq:Om-basis}
   \eta = {\cal E}^T
      \begin{pmatrix} 0  & \mathbb{I} \\  \mathbb{I} & 0  \end{pmatrix} {\cal E} ,
      \qquad
   {\cal H} = {\cal E} ^T
      \begin{pmatrix} \mathbb{I} & 0 \\ 0 &  \mathbb{I}
\end{pmatrix} {\cal E} .
\eeq

As already discussed, we will be interested in  solutions  where the manifold $M$ is a $n$-dimensional torus fibration
(with coordinates $y^m$) over a  base $\mathcal{B}$ (with coordinates $x^{\mu}$)
\beq
\label{eq:fibermetric}
{\rm d} s^2=  g_{\mu\nu} {\rm d}x^{\mu} {\rm d}x^{\nu} + g_{m n} ( {\rm d}y^m + A^{m}_{\ \ \rho} {\rm d}x^{\rho} ) 
( {\rm d}y^n+A^n_{\ \ \sigma}{\rm d}x^{\sigma} ) \, .
\eeq
The corresponding vielbeins are
\bea
e^\alpha&=&e^\alpha_{\ \ \mu} {\rm d}x^{\mu} \\
e^a&=&e^a_{\ \ m} ( {\rm d}y^m + A^m_{\ \ \nu} {\rm d}x^{\nu} ) = e^a_{\ \ m} \Theta^m \, ,
\eea
where $\alpha$ and $a$ are the local Lorentz indices on the base and the fibre,
respectively, while $\mu$ and $m$ are the corresponding target-space indices.
We take also a non trivial $B$-field of the form
\bea
B &=& B^{(2)} + B^{(1)} + B^{(0)} \nn \\
&=& \frac{1}{2} B_{\mu \nu} \,   {\rm d}x^{\mu} \wedge {\rm d}x^{\nu}
+  B_{\mu m} \, {\rm d}x^{\mu} \wedge \Theta^m  +
\frac{1}{2} B_{mn} \,  \Theta^m  \wedge  \Theta^n  \, ,
\eea
where $B^{(2)}$ is the component entirely on the base, $B^{(1)}$ has one component on the base and one on the
fibre, and $B^{(0)}$ is on the fibre
\bea
B^{(2)} &=& \frac{1}{2} (B_{\mu \nu} - 2 B_{m [\mu} A^{m}_{\ \ \nu]} + B_{mn}A^{m}_{\ \ \mu}A^{n}_{\ \ \nu} )\ 
{\rm d}x^{\mu}\w {\rm d}x^{\nu} \, , \\
B^{(1)} &=& ( B_{\mu m} - B_{mn}A^{n}_{\ \ \mu})\ {\rm d}x^{\mu}\w {\rm d}y^{m}  \, , \\
B^{(0)} &=&  \frac{1}{2} B_{mn}\ {\rm d}y^{m} \w {\rm d}y^{n} \, .
\eea

The generalized vielbeins in \eqref{eq:Om-basis} then take the form
\begin{equation}
\label{eq:genvfiber0}
\cE^A{}_M {\rm d}X^M = \begin{pmatrix} e & 0 \\ -\hat{e} B & \hat{e} \end{pmatrix} \,
\begin{pmatrix} {\rm d}x \\ \del \end{pmatrix}
= \begin{pmatrix}
e^{\alpha}{}_{\mu} & 0 & 0 & 0   \\
A^{a}{}_{\mu} & e^{a}{}_{m} & 0 & 0 \\
- B_{\alpha \mu}  &  - B_{\alpha m}  &  {\hat e}_{\alpha}{}^{\mu} &
\hat{A}_{\alpha}{}^{m}  \\
- B_{a \mu}  &  - B_{a m} & 0 & \hat{e}_{a}{}^{m}
 \end{pmatrix}  \, \begin{pmatrix} {\rm d}x^{\mu}\\ {\rm d}y^{m}
\\  \partial_\mu \\\partial_{m} \end{pmatrix}  \, ,
\end{equation}
where $\hat{e} = (e^{-1})^T$.
To simplify the notation  we defined the connections
$A^{a}{}_\nu = e^{a}{}_{m} A^{m}{}_\nu$ and
$\hat{A}_{\alpha}{}^{m} = - \hat{e}_{\alpha}{}^{\mu} A_\mu{}^m$.
Similarly the components of the $B$-field are 
\bea
&& B_{\alpha n} = {\hat e}_{\alpha}{}^{\mu} B_{\mu n} \qquad \quad 
B_{\alpha \nu} = {\hat e}_{\alpha}{}^{\mu}  (B_{\mu \nu} + B_{\mu m} A^{m}{}_\nu
- A_\mu{}^m B_{m \nu})  \, ,\\
&&  B_{a n} = {\hat e}_{\alpha}{}^m B_{m n} \qquad \ \  B_{a \nu}  =  {\hat e}_{a}{}^{m} ( B_{m n} A^n{}_\nu + B_{n \nu})  \, . 
\eea

Expression \eqref{eq:genmetric} is well known from the study of T-duality, where it parametrises
the moduli of $d$-dimensional toroidal compactifications, and indeed its transformation under
$O(d,d)$ is the same as in standard T-duality
\beq
\mathcal{H} \mapsto \mathcal{H'} = O^T \mathcal{H} O \, .
\eeq

Then the $O(d,d)$ transformations of the generalized vielbeins follow immediately
\beq
\cE \mapsto \cE' = \cE O \, .
\eeq
Note that the choice of generalized vielbeins \eqref{eq:genvfiber0} is
invariant under the $\Ggeom$ subgroup of $O(d,d)$ transformations. \\

As an example of the transformation \eqref{eq:O(6,6)tr},
consider now a manifold which is a direct product of a base and a ``fibre'' and with no $B$-field. The generalized vielbeins take the simple form
\beq
\cE= \begin{pmatrix} e_{\bbb} & 0 & 0 & 0  \\ 0 & e_{\fff} & 0 & 0 \\
0 & 0 & \hat{e}_{\bbb} & 0 \\ 0 & 0 & 0 & \hat{e}_{\fff}
\end{pmatrix} \, ,
\eeq
where with obvious notation $e_{\mathcal B}$ and $e_{\mathcal F}$ denote the vielbeins on
the base and the fibre.
After the transformation \eqref{eq:O(6,6)tr}, it becomes
\beq
\cE' = \begin{pmatrix} e_{\bbb} A_{\bbb} & 0 & 0 & 0  \\ e_{\fff} A_{\ccc} & e_{\fff} A_{\fff} & 0 & 0 \\
\hat{e}_{\bbb} C_{\bbb} & \hat{e}_{\bbb} C_{\ccc} & \hat{e}_{\bbb} D_{\bbb} & \hat{e}_{\bbb} D_{\ccc} \\ \hat{e}_{\fff} C_{\ccc '}& \hat{e}_{\fff} C_{\fff} & 0 & \hat{e}_{\fff} D_{\fff}
\end{pmatrix} \, .
\eeq
Comparing the previous expression with \eqref{eq:genvfiber0}, it is easy to see that the new background has a non-trivial $B$-field
\beq
\label{eq:Bfield}
B^\prime =-A^T C= - \begin{pmatrix}
\tilde{C}_{\bbb} &  - \tilde{C}^T_{\ccc}  \\ \tilde{C}_{\ccc }& \tilde{C}_{\fff}
\end{pmatrix} \, ,
\eeq
and a non-trivial fibration structure with connection 
$A^\prime = A_\fff^{-1} A_{\ccc}$.
The transformed metric is then
\beq
\label{eq:newmet}
{\rm d}s^2=  g^\prime_{\mu\nu} {\rm d}x^{\mu} {\rm d}x^{\nu} + g^\prime_{m n} 
( {\rm d}y^m + A^{\prime \ \!\! m}_{\ \ \rho}{\rm d}x^{\rho} )
( {\rm d}y^n + A^{\prime \ \!\! n}_{\ \ \sigma}{\rm d}x^{\sigma} ) \, ,
\eeq
where $g^\prime_{\mu \nu} =  (A^T_{\bbb} \, g_{\bbb} \, A_{\bbb})_{\mu\nu}$ and $g^\prime_{m n} =  (A^T_{\fff} \, g_{\fff} \, A_{\fff})_{m n}$.
Similarly, from the usual $O(d,d)$ transformation of the dilaton we get
\beq
\label{dil0}
e^{\phi^\prime}=e^{\phi} \left[ \frac{\det(g^\prime)}{\det(g)} \right]^{\frac{1}{4}} \, .
\eeq
and from the explicit form of the metrics $g$ and $g'$, \eqref{eq:newmet}, we have
\beq
\label{dil1}
e^{\phi^\prime}=e^{\phi} |\det(A_{\mathcal{B}})\ \det(A_{\mathcal{F}})|^{\frac{1}{2}} \ .
\eeq

The matrices $A_\bbb$, $A_\fff$, $A_\ccc$, $\tilde{C}_{\bbb}$, $\tilde{C}_{\fff}$, and $\tilde{C}_{\ccc}$  are completely arbitrary, and hence the transformation \eqref{eq:O(6,6)tr} allows to go from whatever metric, dilaton  and $B$-field, to any other metric, dilaton, connection, and $B$-field.

\subsection{Action of the transformation on pure spinors}

We would like to use our twist transformation to generate new solutions. Since we are dealing with supersymmetric compactifications, we can concentrate on solving the supersymmetry conditions\footnote{If there are non trivial fluxes, the Bianchi identities for the fluxes must also be imposed.}, since they are equivalent to the full system of equations of motion\cite{LT,KT}. The supersymmetry variations in type II supergravity can be expressed in the language of Generalized Geometry as differential equations on a pair of compatible $O(6,6)$ pure spinors \cite{GMPT04,GMPT05}. We will discuss the SUSY equations and their transformations under $O(6,6)$ in the next sections. Here we will focus on a basic ingredient, namely the transformations of the pure spinors under $O(d,d)$. \\

Spinors on $E$ are Majorana--Weyl $\Spin(d,d)$ spinors.  The spin bundle splits into
two chiralities, $S(E)_\pm$ and, in each representation, one can select a vacuum of Cliff$(d,d)$. This defines 
a pure spinor. There is an
isomorphism between pure spinors and even/odd forms on $E$ 
\begin{equation}
\label{eq:PhiL}
   \Psi_\pm \in L \otimes
      \Lambda^{\text{even/odd}}T^*M  \, .
\end{equation}
$L$ is a trivial line bundle which, as explained in \cite{GMPW}, essentially reflects the presence of the 
dilaton in the pure spinor: the  
sections of $L$ are given by $e^{-\phi}$. Note that the isomorphism \eqref{eq:PhiL} is defined up to a 
multiplication by a complex number and, in general, defines line bundles of pure spinors.
On a symplectic manifold,  this line bundle is generated by the 
exponential of the  symplectic form and can always be trivialized.  In the case
of a complex manifold, $L$ is the usual canonical line bundle. This means, in particular, that we can fix the 
phase and have global pure spinors 
only on a manifold with a vanishing first Chern class. In general this condition is not satisfied. For some 
of our applications the phase is important and hence we shall keep it explicit.  However, by a slight abuse of 
language, we will refer to the lines of pure spinors as simply pure spinors.

Two pure spinors are said to be
compatible when they have $d/2$ common annihilators. Two compatible pure spinors
define an $U(d) \times U(d)$ structure on $E$ (the structure group is reduced to $SU(d) \times SU(d)$ when 
the line bundles of the 
complex differential forms can be trivialized).
Any pure spinor can be represented as a wedge product of an exponentiated complex
two-form with a complex $k$-form. The degree $k$ of the form is the type of the pure spinor.
The explicit expression for a pair of compatible
pure spinors depends on
the geometry of the manifold $M$. For instance, if $M$ has $SU(3)$ structure, 
two compatible pure spinors are of type 0 and type 3 
\bea
\label{eq:Su3purespin1}
&& \Psi_+=  e^{i\theta_+}  e^{-\phi} e^{-B} \, e^{-iJ} \, , \\
\label{eq:Su3purespin2}
&& \Psi_-= -i e^{i \theta_-}  e^{-\phi} e^{-B} \Omega \, ,
\eea
with $J$  the real K\"ahler form and $\Omega$ the holomorphic three-form on $M$.\\

The $O(d,d)$  action on a spinor on $E$ is constructed in the usual way.
We define the $O(d,d)$ generators in the spinorial representation as
\beq
\sigma^{MN}=[\Gamma^M,\Gamma^N] \, ,
\eeq
with $M,N$ $d+d$ indices. 
Then the group element in the spinorial representation is
\beq
\label{O(d,d)sp}
O  = e^{-\frac{1}{4}\Theta_{MN} \sigma^{MN}} \, ,
\eeq
and it acts on the spinors  by wedges and contractions: $\Gamma^n = {\rm d}x^n \w$ and $\Gamma_m = \iota_{\del_m}$.
The matrix $\Theta_{M N}$ is antisymmetric and reads
\beq
\Theta_{M N} = \begin{pmatrix}
a^m_{\ \ n} & \beta^{m n} \\
B_{m n} & - a_m^{\ \ n}
\end{pmatrix} \, ,
\eeq
where $a^m_{\ \ n}$, $B_{m n}$ and $\beta^{m n}$ parametrise the generators of the $GL(d)$
transformations, $B$-transform and $\beta$-transform, respectively. 
In the next sections, we will need 
the explicit action of $GL(d)$ and  $B$-transform, \cite{G}.
The $GL(d)$ action is given by 
\bea
\label{GL(d)sp}
O_a  
&=& e^{- \frac{1}{4}(a^m_{\ \ n} [\Gamma_m,\Gamma^n]-a_m^{\ \ n}[\Gamma^m,\Gamma_n])} \nn \\
&=& e^{-\frac{1}{2} Tr(a) +a^m_{\ \ n} {\rm d}x^n \wedge \, \iota_{\del_m}} \nn \\
&=& \frac{1}{\sqrt{\det A}} \, \,  e^{a^m_{\ \ n} {\rm d}x^n \wedge \, \iota_{\del_m}}  \, .
\eea
Similarly, for a $B$-transform, we obtain
\bea
\label{Btrsp}
O_B  
&=& e^{- \frac{1}{2} B_{m n}\Gamma^{m n}}  \nn \\
&=& e^{-\frac{1}{2} B_{m n} {\rm d}x^m \wedge {\rm d}x^n } \, .
\eea

Given the action of the twist transformation on the generalized vectors, we want to know what
is the corresponding transformation on the pure spinors. To make the
exponentiation easier, we can decompose the matrix \eqref{eq:O(6,6)tr} as a product
\beq
\label{3prod}
O = \begin{pmatrix} A & 0  \\ C & A^{-T} \end{pmatrix} 
= \begin{pmatrix} \mathbb{I} & 0  \\ X & \mathbb{I} \end{pmatrix}
\begin{pmatrix} A & 0 \\ 0 & A^{-T} \end{pmatrix}
\begin{pmatrix} \mathbb{I} & 0 \\ - Y  & \mathbb{I} \end{pmatrix}
\eeq
with $Y = A^T X A - A^T C$. In the transformation of the generalized vielbein, we showed
that the $B$-field of the transformed background is $B^\prime=A^T B A -A^T C$.
Therefore we can interpret $X$ as the $B$-field of the original solution and $Y$ as
the new one.
Similarly, \eqref{3prod} can be seen as a succession of a $B$-transform, a
$GL(d)$ rotation and another $B$-transform.
This leads to the following expression for the $O(6,6)$ action on the spinors
\beq
\label{Otrsp}
O_f  = \frac{1}{\sqrt{\det A}} e^{- y_{m n} {\rm d}x^m \wedge {\rm d}x^n}
\, e^{a^m_{\ \ n} {\rm d}x^n \wedge \, \iota_{\del_m}}  \,
e^{x_{m n} {\rm d}x^m \wedge {\rm d}x^n} \ .
\eeq 
Since $O(6,6)$  acts  on the generalized 
vielbein from the right and on pure spinors from the left, we have exchanged the
order of the transformations with respect to \eqref{3prod}. \\

Finally, when applying the transformation  \eqref{eq:O(6,6)tr} to the pure spinors we should allow for an arbitrary phase. 
This reflects the freedom to change section of the line bundle $L$.
Hence,  we should add to the action of (\ref{3prod}) a $U(1)$ shift by $\mbox{exp}(i \theta_c^{\pm})$:
\beq 
\label{gentwist}
O_c^{\pm} = e^{i\theta_c^{\pm}} O_f \, .
\eeq

\subsubsection{Twisting $\mathbb{T}^2$}

In this section we apply the twist transformation to our main examples, manifolds $M$ which are 
$\mathbb{T}^2$ fibration over a four-dimensional base $\mathcal{B}$
\beq 
\bbb \times \mathbb{T}^2 \Rightarrow   \mathbb{T}^2\hookrightarrow M \stackrel{\pi} \longrightarrow {\bbb} \, .
\eeq
Depending on the example, we take $\bbb$ to be $\mathbb{T}^4$, or $K3$. We will denote the holomorphic coordinate on the fibre by $z = \theta^1 + i \theta^2$. Then
the torus generators are defined as  $\partial_{z}$ and $\partial_{\bar z}$,
and the connection  one-forms by $\Theta^I = {\rm d} \theta^I + A^I$, with $\Theta=\Theta^1 + i\Theta^2$ and $\alpha = A^1 + iA^2$. 
The fibration will be in general non-trivial, 
and the curvature two-forms $F^I \in \Omega^2_{\bbZ}({\cal B})$ are given by ${\rm d} \Theta^I = \pi^* F^I$.\\

Our starting point is a trivial $\mathbb{T}^2$ fibration.
For simplicity, we set the $B$-field to zero
and the dilaton to a  constant. The pure spinors are as in
\eqref{eq:Su3purespin1} and  \eqref{eq:Su3purespin2}, with the SU(3) structure defined by
\bea
&& J = J_{\bbb} + \frac{i}{2} g_{z\ov{z}} {\rm d}z \wedge {\rm d}\bar{z} \\
&& \Omega = \sqrt{g} \, \omega_{\bbb} \wedge {\rm d}z \, ,
\eea
where $g$ is the determinant of the
metric on the torus fibre,
$J_{\bbb}$  and $\omega_{\bbb}$ the K\"ahler and  holomorphic
two-forms on the base.\\

In the transformation \eqref{Otrsp} we set $x_{mn}=0$ since there is no initial $B$-field,
and take $y_{mn}$ an arbitrary antisymmetric matrix. This will act
as a standard $B$-transform  giving the new $B$-field. Here we will concentrate
on the $GL(6)$ part. For simplicity, we take the action on the base to be trivial 
\beq
A= \begin{pmatrix} 1_4 & 0 \\ A_{\ccc} & A_{\fff} \end{pmatrix} = 
\begin{pmatrix} 1_4  & 0 \\ 0 & A_{\fff} \end{pmatrix}
\begin{pmatrix} 1_4 & 0 \\  A^\prime  & 1_2 \end{pmatrix}
\eeq
and 
\beq
A_{\fff} = \begin{pmatrix} e^{\lambda_1} & 0 \\ 0  & e^{\lambda_2} \end{pmatrix}
\qquad   A^\prime =A_{\fff}^{-1} A_{\ccc} = \begin{pmatrix} A^1_{\ \ \mu}  \\ A^2_{\ \ \nu}  \end{pmatrix} \, .
\eeq
With this choice, the $GL(6)$ factor in \eqref{Otrsp} becomes
\beq
\label{OtrspT2}
O_a = \frac{1}{\sqrt{\det A_{\fff}}} 
\, e^{A^1  \, \iota_{\del_1} + A^2  \, \iota_{\del_2}}   \,
e^{\lambda_1  \, {\rm d}x^1\w \, \iota_{\del_1} + \lambda_2  {\rm d}x^2 \w \, \iota_{\del_2}} \, ,
\eeq 
with $A^I = A^I_{\ \ \mu}  {\rm d}x^\mu$ for $I=1,2$.
In terms of the complex connection
$\alpha$, the off-diagonal block becomes 
\bea
\label{tracelessa}
&& A^I  \, \iota_{\del_I} =\alpha \w i_{\partial z} + \overline{\alpha} \w i_{\partial \overline{z}} \nn \\
&& e^{A^I  \, \iota_{\del_I}}=1+(\alpha \w i_{\partial z} + \overline{\alpha} \w i_{\partial \overline{z}})+\alpha \w \overline{\alpha} \w i_{\partial \overline{z}}\ i_{\partial z}=1+o \cdot \label{otr} \, ,
\eea
where $o \cdot$ sends a form to another form with same degree. The diagonal blocks give
\beq
\label{tracelessb}
e^{\lambda_{I} {\rm d}x^I \wedge \, \iota_{\del_I}}
= \prod_{I=1,2} \left[ 1+ (e^{\lambda_I} -1 ) {\rm d}x^I \wedge \, \iota_{\del_I} \right] \, .
\eeq
To derive \eqref{tracelessb} we used the fact the operators ${\rm d}x^I \wedge \, \iota_{\del_I}$
commute for different values of $I$,  and $({\rm d}x^I \wedge \, \iota_{\del_I})^k = {\rm d}x^I \wedge \, \iota_{\del_I}$.\\

The effect of \eqref{tracelessb} on $\Omega$ and $J$ is to rescale the fibre components, while \eqref{tracelessa}
introduces the shift of the fibre direction by the connections $\alpha$ and $\overline{\alpha}$
\bea
\label{SU(3)strnew}
&& J^\prime =J_{\mathcal{B}}+ \frac{i}{2} g^\prime_{z\ov{z}}  \, \Theta \w  \overline{\Theta} \\
&& \Omega^\prime =  \sqrt{g^\prime} \,  \omega_{\mathcal{B}} \w \Theta \, , 
\eea
where $g^\prime_{z\ov{z}} = e^{2 \lambda} g_{z\ov{z}}$ and $\Theta = {\rm d}z+\alpha $.
In order not to change the complex structure, we have to
set $\lambda_1 = \lambda_2 = \lambda$.\\

Finally, from \eqref{tracelessa} and \eqref{tracelessb}, it is straightforward to compute
the new pure spinors
\bea \label{newpsT2}
&& \Psi_+ =  e^{i\theta_+} e^{-\phi}\, e^{-iJ} \quad \longrightarrow \quad
\Psi_+^\prime= e^{i\theta_+} e^{-\phi'} e^{-B^\prime} e^{-iJ^\prime} \ , \nn\\
&&\Psi_- = -i  e^{i \theta_-} e^{-\phi}\, \Omega \,  \quad \longrightarrow \quad 
\Psi_-^\prime= - i e^{i \theta_-} e^{-\phi'} e^{-B^\prime} \Omega^\prime \, .
\eea
Here we took the normalized pure spinors and we did not transform the phases.
The new $B$-field is clearly $B' = y_{m n} {\rm d}x^m \wedge {\rm d}x^n$ and the dilaton
is transformed by the trace part of the $GL(6)$ transformation
\beq
\label{newdil}
e^{- \phi^\prime} = ( \det A_{\fff})^{-1/2} e^{-\phi} = e^{-\lambda} e^{-\phi}\, .
\eeq

\subsubsection{$SU(2)$ structures}

In all our examples the base manifold $\bbb$ is a complex manifold and we can write the metric on $M$ as 
\beq\label{metr}
{\rm d}s_M^2 = e^{2\Delta} g_{i {\bar j}} {\rm d}z^i {\rm d}\bar{z}^j + g_{z \ov{z}}\, \Theta {\overline \Theta} \, ,
\eeq
where $g_{i{\bar j}}$ and ${\rm d}z^i$ ($i=1,2$) are the metric and complex  coordinates on the base manifold $\bbb$, 
respectively.
In the examples $\Delta$ can be related to the dilaton $\phi$. The associated K\"ahler form and holomorphic three-form 
are given in terms of the base two-forms, $J_{\mathcal{B}}$ and $\omega_{\mathcal{B}}$, and
and the fibre one-form as in  \eqref{SU(3)strnew}. 

Note that $J_{\mathcal{B}}$ and $\omega_{\mathcal{B}}$ in  \eqref{SU(3)strnew}
define an $SU(2)$ structure on the base manifold $\bbb$. 
We can use the pair of vectors $\partial_{z}$ and $\partial_{\bar z}$ to define a local $SU(2)$ structure also on the whole manifold $M$. 
This can be done in terms of  a complex 1-form $Z$,  a real 2-form $j$ and a 
complex (2,0)-form $\omega$ satisfying
\bea \label{su2str}
&& \omega \wedge j=\omega \wedge \omega =0 \, , \nonumber \\
&& j^2 = \frac{1}{2} \omega \wedge \bar{\omega} \, , \nonumber \\
&& Z \llcorner j = Z \llcorner \omega = 0  \, .
\eea
The one-form $Z$ is dual to the complexified vector field. To define the two-forms we shall use an alternative parametrisation of the metric (\ref{metr})
\beq\label{metr2}
{\rm d}s_M^2 =  {\tilde g}_{i {\bar j}} \chi^i \bar{\chi}^{\bar j} + Z  {\overline Z} \, ,
\eeq
where 
\bea
{\tilde g}_{i {\bar j}}  &=& e^{2\Delta} g_{i {\bar j}}  + g_{z \ov{z}} A_i A_{\bar j} \, , \nn \\ 
\chi^i &=& {\rm d}z^i + g_{z \ov{z}} \, \tilde{g}^{i {\bar j}} A_{\ov{j}} \, {\rm d} z \, , \nn\\
Z &=& \sqrt{g_{z \ov{z}} - g_{z \ov{z}}^2 \, \tilde{g}^{i {\bar j}} A_i A_{\bar j}}  \, {\rm d} z \, .
 \eea
The $SU(2)$ structure, or alternatively a pair of $SU(3)$ structures, on $M$ is given now by 
\bea
\label{SU(2)strneww}
&& J_{\pm} = \frac{i }{2} \, {\tilde g}_{i {\bar j}} \chi^i \wedge \bar{\chi}^{\ov{j}}
  \pm \frac{i }{2} {Z \wedge \ov{Z}} \, , \nn \\ 
&& \Omega = \sqrt{\det {\tilde g}} \, \chi^1 \wedge \chi^2 \wedge Z_{\pm} \, ,
\eea 
where $Z_+ = Z$ and $Z_- = \ov{Z}$. Using $j$, $\omega$ and $Z$, more general pure spinors of mixed type 0 - 
type 1 can be constructed. See for instance \cite{MPZ, KT, An} for such solutions.

\section{Type II transformations}
\label{tpII}

One of our aims is to apply the $O(6,6)$ transformation discussed
in the previous section as a solution
generating technique in Type II string theory.
The idea would be to start from a known solution and get a new one. Instead of
studying when \eqref{eq:O(6,6)tr} preserves the equations of motion, we will
focus on the supersymmetry variations and derive the conditions that
supersymmetry imposes on the $O(6,6)$ transformation
in order for the latter to map solutions to new ones.
We would like to stress that, in this paper, we do not solve the general conditions
the $O(6,6)$ transformation has to satisfy. We will leave it to future work.
Here we concentrate on an explicit application of our transformation
to the context of SU(3)
structure compactifications on $\mathbb{T}^6$ or six-dimensional twisted tori.

\subsection{Generating solutions: constraints in type II and RR fields transformations}

We will be interested in type II backgrounds corresponding to warp products
of four-dimensional Minkowski times  a six-dimensional compact manifold. 
The ten-dimensional metric in string frame is given by
\beq
\label{metricansatz}
{\rm d}s_{(10)}^2=e^{2A(y)}\ \eta_{\mu\nu} {\rm d}x^\mu {\rm d}x^\nu +g_{m n}(y) {\rm d}y^m {\rm d}y^n \ ,
\eeq
where $\eta$ is the diagonal Minkowski metric. \\

The supersymmetry conditions for Type II compactifications have been given in
the language of Generalized Geometry in \cite{GMPT04,GMPT05} as a set of differential equations for 
a pair of compatible pure
spinors on $E$. In $\mathcal{N}=1$ compactifications the supersymmetry parameters decompose as
\beq
\epsilon^i = \zeta_+ \otimes \eta^i_+ + \zeta_- \otimes \eta^i_- \qquad i=1,2 \, ,
\eeq
for type IIB, while for type IIA
\beq
\epsilon^1 = \zeta_+ \otimes \eta^1_+ + \zeta_- \otimes \eta^1_- \qquad 
\epsilon^2 = \zeta_+ \otimes \eta^2_- + \zeta_- \otimes \eta^2_+ \, .
\eeq
In both cases $\zeta_+$ is a 4$d$ Weyl spinor of positive chirality ($\zeta_- = (\zeta_+)^\ast$) and $\eta^i_+$ are
two positive chirality spinors in six dimensions ($\eta^i_- = (\eta^i_+)^\ast $). By tensoring the internal spinors $\eta^1$ and
$\eta^2$, we construct a pair of $O(6,6)$ spinors\footnote{There is an isomorphism between $O(6,6)$ spinors and bispinors (tensor product of Cliff$(d)$ spinors)
given by the Clifford map
\beq
\label{eq:cliffmap}
C = \sum_{p} \frac{1}{p!} C_{i_1 \ldots i_p} {\rm d}x^{i_1} \wedge \ldots
\wedge  {\rm d}x^{i_k} \qquad \leftrightarrow \qquad
C_{\alpha \beta} = \sum_{p} \frac{1}{p!}C_{i_1 \ldots i_p} \gamma_{\alpha \beta}^{i_1 \ldots i_k}  \, ,
\eeq
where $\gamma^{i_k}$ are $2^{d/2}\times 2^{d/2}$ matrices.}
\beq
\label{purespina}
\Phi_\pm = \eta^1_+ \otimes \eta^{2 \, \dagger}_\pm \, .
\eeq
These are by construction pure and compatible, and define a $SU(3) \times SU(3)$ structure on $E$.

The explicit form of the pure spinors $\Phi_\pm$ depends on the relation between the
two supersymmetry parameters $\eta^1$ and $\eta^2$. In this paper we will
mostly consider the case of manifold of $SU(3)$ structure, where there is
a single globally defined spinor $\eta_+$ (with unitary norm). Hence
\bea
\label{eq:Su3sp}
&& \eta^1_+ = |a| \, e^{i \alpha} \eta_+  \, \nn \\
&& \eta^2_+ = |b| \, e^{i \beta} \eta_+ \, .
\eea
Here $|a|$ and $|b|$ are the norms of $\eta^{1,2}$. Supersymmetry sets them equal and proportional to the warp 
factor:
$|a|= |b|= e^{A/2}$. The corresponding pure spinors take the form \eqref{eq:Su3purespin1} and \eqref{eq:Su3purespin2}
\bea
\Psi_+ &=& 8 \ \ee^{-\phi}\ee^{-B} \frac{\Phi_+}{||\Phi_+||} \nn \\
&=& e^{i\theta_+} e^{-\phi} e^{-B} \, e^{-iJ} \, , \label{eq:Psiplus} \\
\Psi_- &=& 8 \ \ee^{-\phi}\ee^{-B} \frac{\Phi_-}{||\Phi_-||}  \nn \\
&=& -i e^{i\theta_-} e^{-\phi} e^{-B} \Omega \, ,
\eea
where $\theta_+ = \alpha - \beta$, $\theta_-  = \alpha + \beta$, and $||\Phi_{\pm}||=|a|^2$.
$J$ is the real K\"ahler form and $\Omega$ the holomorphic three-form on $M$. We choose to work with twisted normalised pure spinors since they are
those transforming nicely under $O(6,6)$. \\

The supersymmetry variations for the gravitinos and
dilatinos are completely equivalent to the following differential conditions on the pure spinors
\bea
\label{eq:susyeqII}
&& {\rm d}(e^{3A}\Psi_1)= 0 \, , \nn\\
&& {\rm d}(e^{2A} \re \Psi_2) =  0 \, ,\nn\\
&& {\rm d}(e^{4A} \im\Psi_2) =  e^{4A} e^{-B} \ast \lambda(F)  \, .
\eea
Thus in the Generalized Complex Geometry language a necessary condition for ${\cal N}=1$ supersymmetric backgrounds is
to have a twisted Generalized Calabi-Yau manifold:
one pure spinor must be $H$-closed.
In type IIA the closed pure spinor is the even one, $\Psi_1= \Psi_+$, while
in type IIB it is the odd one $\Psi_1=\Psi_-$.
The non integrability of the second spinor  is due to the RR fluxes on the internal manifold. In \eqref{eq:susyeqII} we denote by $F$ the sum of the RR fluxes on the internal manifold
\bea
\textrm{IIA}&:&\ F=F_0+F_2+F_4+F_6 \ , \\
\textrm{IIB}&:&\ F=F_1+F_3+F_5 \ .
\eea
$F$ is related to the total ten-dimensional
RR field-strength $F^{(10)}$ by
\beq
F^{(10)}=F+\textrm{vol}_{(4)}\w \lambda(*F) \ ,
\eeq
where $\textrm{vol}_{(4)}$ is the warped four-dimensional volume form with warp factor
 $e^{2A}$. Finally, $\lambda$ acts on any $p$-form $A_p$  as the complete reversal of its indices
\beq
\lambda(A_p)=(-1)^{\frac{p(p-1)}{2}}A_p \label{lambda} \ .
\eeq


Supersymmetry only sets necessary conditions for $\mathcal{N}=1$ vacua.
In order to have a full solution, the Bianchi identities for the fluxes must be
imposed\footnote{It has been proven \cite{GMPT06, KT} that the equations of motion
are implied by supersymmetry and the Bianchi identities.}
\beq
({\rm d} - H \w ) F=\delta(source) \ , \qquad {\rm d}H=0 \ . \label{BI}
\eeq
Here $\delta(source)$ is the charge density of the sources: these are space-filling
D-branes or orientifold planes (O-planes).\\

Consider now a solution of the supersymmetry equations and Bianchi identities,
\eqref{eq:susyeqII} and \eqref{BI}, and apply to the associated pure spinors the transformation \eqref{gentwist}
\beq
\label{compO(6,6)}
O^\pm_c = e^{ i \theta_c^\pm} O_{f} \qquad \Rightarrow \qquad  \Psi_\pm^\prime =
O^\pm_c  \,  \Psi_\pm \, .
\eeq
We want to determine
what are the conditions on $O_c$ in order to get a new solution. 
Since the existence of a closed pure spinor is a necessary condition for preserving supersymmetry, the idea is to consider transformations
of the form (\ref{gentwist}) that preserve the closure of at least one pure spinor and hence  at least ${\cal N}=1$ supersymmetry.
The action of the transformation on the rest of the fields is then used to define the transformed RR fields.

The condition to get new solutions are easily
determined by imposing that the transformed pure spinors are again
solutions of the SUSY equations
\bea
\label{eq:IInew}
{\rm d}(e^{3A} \Psi_1^{\prime})&=&0 \nn\\
{\rm d}(e^{2A} \textrm{Re}\Psi_2^{\prime})&=& 0\nn\\
{\rm d}(e^{4A} \textrm{Im}\Psi_2^{\prime})&=& R^{\prime} \, ,
\eea
where $R^{\prime}$ is the new RR field ($R= e^{4A} e^{-B} \ast \lambda(F)$).  Then expanding
into real and imaginary parts, we obtain
\bea
\label{cont0}
&& {\rm d}(O_f) \, \Psi_1 = 0 \nn\\
&& \cos \theta_c^+ \,  {\rm d} (O_f) \,  e^{2 A} \re \Psi_2 - \sin \theta_c^+ \,  {\rm d} (e^{-2A} O_f) \,  
e^{4 A} \im \Psi_2
= e^{- 2A} \sin\theta_c^+ \, O_f \,  R \nn\\
&& \sin \theta_c^+ \,  {\rm d} (e^{2 A} O_f) \,  e^{2A}  \re \Psi_2 + \cos \theta_c^+ \,  {\rm d} (O_f) \,  
e^{4A} \im \Psi_2 =
R^{\prime} - \cos \theta_c^+ \,  O_f \,   R \ .
\eea
The last equation defines the transformed RR field
\beq
\label{RRprime}
R^{\prime}=\cos(\theta_c^+) O_f \,  R + \sin(\theta_c^+) {\rm d} (e^{2A} O_f) \,   e^{2A} \re \Psi_2 + \cos(\theta_c^+)
{\rm d} (O_f) \,  e^{4A} \im \Psi_2 \ .
\eeq
The first two are the constraints the $O_c$ transformation has to fulfill in order
to map solutions to new solutions of type II supergravity.
As already mentioned at the beginning of this section, we do not analyse in general
the system of constraints above.

An interesting feature of this transformation is the possible mixing between the NSNS
and RR sectors. This is due to the complexification of the $O(6,6)$ transformation
by the $U(1)$ action on the line of pure spinors. Note also that such a complexification is necessary
to relate different types of sources.

\subsection{Mapping solutions in torus compactifications}
\label{map}

A relatively simple and still non trivial class of flux compactifications are provided by T-duals of toroidal compactifications\cite{KSTT, Sc}. 
In all these cases the ten-dimensional metric is of
the form \eqref{metricansatz}, where the six-dimensional manifold can be the straight $\mathbb{T}^6$ or a twisted torus.
Such manifolds also provide explicit examples of Generalized Calabi-Yau manifolds \cite{CG,GMPT06}. \\

In this section we will use our twist transformation \eqref{eq:O(6,6)tr} to relate compactifications on $\mathbb{T}^6$ to nilmanifolds that are fibrations 
of $\mathbb{T}^2$ over $\mathbb{T}^4$
\beq 
\mathbb{T}^4 \times \mathbb{T}^2 \Rightarrow   \mathbb{T}^2\hookrightarrow M \stackrel{\pi} \longrightarrow {\mathbb{T}^4} \, .
\eeq   
As in the previous section, we will denote the torus generators  by $\partial_{z}$ and $\partial_{\bar z}$,
and the connection  one-forms by $\Theta^I = {\rm d} \theta^I + A^I$ 
and the curvature two-forms by $F^I$ with $I=1,2$.\\

The twist transformations  necessarily relate manifolds with different topological properties. This can be seen by computing the Betti numbers of the different
manifolds. For the direct product of $\mathbb{T}^2$ with a generic base $\bbb$, the Betti numbers are  
\bea
\label{bettin}
&& b_1 = b_1(\bbb) + 2 \ , \nn\\
&& b_2 = b_2(\bbb) + 2 b_1(\bbb) +1 \, , \nn\\
&& b_3 = b_3(\bbb) + 2 b_2(\bbb) + b_1(\bbb) \, .
\eea
Clearly the Betti numbers for generic $M$ are smaller than for $\bbb \times \mathbb{T}^2$ and will depend on the topological properties of the curvature $F$.   
Indeed as ${\rm d}\theta^I$ is mapped to $\Theta^I = {\rm d}\theta^I + A^I$  and ${\rm d} \Theta^I = \pi^* F^I$, 
the two one-forms ${\rm d}\theta^I$, which were non-trivial in cohomology, are replaced by forms that are not closed. 
At the same time the two closed two-forms $F^I$, while being non-trivial in cohomology on $\bbb$, are trivial 
in cohomology on $M$.
When the base is $\mathbb{T}^4$, we find $b_1(M) = b_1(\mathbb{T}^4) = 4$. There are
only seven classes of nilmanifolds with $b_1=4$. 
It is not hard to check that three of them are actually affine $\mathbb{T}^2$ fibrations over $\mathbb{T}^4$ (circle fibrations over five-manifolds which are in turn circle fibrations over $\mathbb{T}^4$)
\begin{center}
\begin{tabular}{lll}
$n$ 4.1 & (0, 0, 0, 0, 12, 15+ 34) &  $M=I_6$ \\
$n$ 4.2 & (0, 0, 0, 0, 12, 15) &  $M=\mathbb{T}^2 \times I_4$ \\
$n$ 4.3 & (0, 0, 0, 0, 12, 14+ 25) & $M=S^1  \times I_5$ 
\end{tabular}
\end{center}
where $I_{n+2}$ is a sequence of two circle fibrations over $\mathbb{T}^n$. This leaves us with four topologically distinct cases of two 
commuting $U(1)$ fibrations\footnote{Note that here we label the nilmanifolds as in 
\cite{GMPT06}, but for $n$ 4.4, $n$ 4.6 and $n$ 4.7
we have used isomorphisms of the nilpotent algebras to cast the individual entries in a 
convenient form, yielding simple
solutions for the same choice of complex structure on the base $\mathbb{T}^4$. The same
isomorphism applied to $n$ 4.5 gives the algebra 
(0,0,0,0, 2 $\times$ (14 - 13) + 23 - 24, 23 - 13 + 2 $\times$ (24 -14)). } over $\mathbb{T}^4$
\begin{center}
\begin{tabular}{lllll}
$n$ 4.5  &  $(0, 0, 0, 0, 12, 34 )$ &  $M= N_3 \times N_3$ &  $b_2(M) = 8$ & $b_3(M)=10$ \\
$n$ 4.6  &  $(0, 0, 0, 0, 13, 14)$ &  $M= S^1 \times N_5$ &  $b_2(M) = 9$ & $b_3(M)=12$ \\
$n$ 4.4  &  $(0, 0, 0, 0, 2\times 13, 14+ 23)$ &  $M= N^{(1)}_6$ & $b_2(M) = 8$ & $b_3(M)=10$ \\
$n$ 4.7  &  $(0, 0, 0, 0, 13 + 42, 14 + 23)$ &  $M= N^{(2)}_6$ &  $b_2(M) = 8$ & $b_3(M)=10$
\end{tabular}
\end{center}
where $N_3$ is a circle fibration over $\mathbb{T}^2$, $N_5$ is a $\mathbb{T}^2$ fibration over $\mathbb{T}^3$ and $N^{(1)}_6$ and $N^{(2)}_6$ are two distinct $\mathbb{T}^2$ fibrations over $\mathbb{T}^4$.\\

Type C solutions, i.e. solutions with a non-trivial RR $F_3$ with O5/D5 sources, can be obtained on some of these 
manifolds by two T-dualities along the fibre from a type B solution on $\mathbb{T}^6$. The latter has a non-trivial five-form 
which is related to the warp factor and an imaginary anti-self dual complex three-form flux $g_s F_3 = - \ast H$.
According to standard Buscher rules, the components of the $B$-field with one leg along the fibre give, after T-duality, the  non-trivial connections. Under T-duality the O3 planes are mapped to O5 planes.

Here we shall show that such manifolds can also be related via our twist transformation \eqref{eq:O(6,6)tr} to
$\mathbb{T}^6$ with O3 planes, a non-trivial five-form flux $F_5$ and a trivial NSNS flux. In this background the five-form flux is related to the warp factor
\beq
g_s F_5=e^{4A} \ast {\rm d}(e^{-4A}) \, ,
\eeq
while the dilaton is constant $e^{\phi}=g_s$. All other fluxes are zero.
The complex structure is chosen as
\bea
\label{T6cs}
&& \chi^1 = e^1+i e^2 \, , \nn\\
&& \chi^2 = e^3+i e^4 \, , \nn\\
&& \chi^3 = e^5+i e^6 \, ,
\eea
where $\chi^i$ are one-forms and the vielbein on the torus are 
$e^i = e^{-A} {\rm d}x^i$ with $i=1, \dots, 6$.
Then the $SU(3)$ structure and the corresponding
pure spinors are
\bea
\label{SU3sol}
\Omega = \chi^1 \wedge \chi^2 \wedge  \chi^3 \, & \qquad \qquad   &
\Psi_- = - \frac{i}{g_s}   \Omega \\
J=\frac{i}{2} \chi^i \wedge \overline{\chi}^i \qquad & \qquad \qquad &
\Psi_+ = \frac{i}{g_s}   \, e^{-iJ} \, .
\eea
The O3 projection fixes one phase $\theta_+= \frac{\pi}{2}$, while we choose for the other $\theta_-=0$.

The idea is now to apply the transformations \eqref{gentwist} and 
\eqref{OtrspT2} to the previous
solution and see under which conditions we can reproduce the nilmanifolds
$n$ 4.5 - $n$ 4.7. We choose the $\mathbb{T}^2$ torus fibre in the directions $x^5$ and
$x^6$. Since we are connecting solutions with zero NSNS flux, we
do not bother to consider the contribution of the $B$-transform.
The new pure spinors are given by \eqref{newpsT2}
\bea
&& \Psi_-^\prime = -i e^{i \theta_c^-} e^{-\phi^{\prime}} \Omega^{\prime} \nn\\
&& \Psi_+^\prime = i e^{i\theta_c^+} e^{-\phi^{\prime}} e^{-iJ^{\prime}} \, ,
\eea
where the $SU(3)$ structure takes the form \eqref{SU(3)strnew} 
\bea
&& J^\prime = J_{\bbb} + \frac{i}{2} g^\prime_{z\ov{z}} \Theta 
\w \ov{\Theta} \\
&& \Omega^\prime =  \sqrt{g^\prime} \,  \omega_{\bbb}
\w \Theta \, ,
\eea
with $J_{\bbb} = \frac{i}{2} ( \chi^1 \wedge \ov{\chi}^{\bar 1} +
\chi^2 \wedge \ov{\chi}^{\bar 2} )$, $\omega_{\bbb} = \chi^1 \wedge \chi^2$
and  $\Theta = {\rm d}z +\alpha$.  

Note that in order to obtain a geometric background, we need to perform 
the twist along isometries. As in standard T-duality, this
implies a smearing in the fibre directions, especially for the warp factor. Then we expect to have 
O5 planes in the directions $56$.

To determine the connection,  as well as the other fields 
in the solution,  we require that the transformed background satisfies 
the supersymmetry constraints \eqref{eq:susyeqII} for O5 compactifications
with type 3 - type 0 pure spinors \cite{GMPT06} 
\bea
&& e^{\phi^\prime} = g_s e^{2A^\prime} \nn \\
&& {\rm d}(e^{A^\prime} \Omega^\prime) = 0 \nn \\
&& {\rm d} (J^\prime)^2  = 0 \nn \\
&& {\rm d}(e^{2A^\prime} J^\prime) = g_s e^{4A^\prime}  *F^\prime_3 \nn\\
&& H =0  \, .
\label{O5cond}
\eea
Also, the O5 projection sets $\theta_+=0$ and we choose again $\theta_-=0$, hence
$\theta_c^- =  0$ and $\theta_c^+ = - \pi/2$.

It is straightforward to verify that from the equation
for the real part of $\Psi^\prime_+$, it follows that indeed 
$e^{\phi^\prime} = g_s e^{2A^\prime}$ and
\beq
\label{rephip}
g^\prime_{z \ov{z}} = e^{2 A^\prime}  \, \qquad \qquad 
\begin{array}{c}
F \w J_{\bbb} = 0 \, , \\
\ov{F} \w J_{\bbb} = 0 \, .
\end{array} 
\eeq
Similarly, the imaginary part of $\Psi^\prime_+$ can be used to define the
RR three-form as in \eqref{O5cond} (see also (\ref{RRprime})).
Finally, the equation for $\Psi^\prime_-$ sets $A^\prime = A$ and
\beq
\label{phim}
F \w \omega_{\bbb} = 0 \, .
\eeq

Using the form \eqref{rephip} for the new metric on the fibre and
the fact that the warp factor does not change, we can write the metric on $M$ as
\beq
\label{O5metr}
{\rm d}s^2_6 =  e^{- 2 A} \sum_{i=1}^{4} ({\rm d}x^i)^2 + 
e^{2 A} \sum_{I=1,2}  ({\rm d}x^I + A^I)^2 \, ,
\eeq
which is indeed what one expects for O5 compactifications.
As a transformation on the generalized vielbein, \eqref{eq:O(6,6)tr}, 
the twist acts as 
\beq
A_{\mathcal{B}}=\mathbb{I}_4 \ , \quad A_{\mathcal{F}}= \mathbb{I}_2 \times e^{2 A^{\prime}} \ , \quad A_{\mathcal{C}\ \ \mu}^{\ \ z}=e^{2 A^{\prime}} \alpha_{\mu}\ , \quad A_{\mathcal{C}\ \ \mu}^{\ \ \ov{z}}=e^{2 A^{\prime}} \ov{\alpha}_{\mu} \ ,
\eeq
and we can check the dilaton is transformed as expected.\\

Let us go back to the form of the constraints on the curvature $F$. From \eqref{rephip} and \eqref{phim}, we see that demanding that the twist preserves supersymmetry is equivalent to the requirement that $F$ does not have a purely anti-holomorphic part and its contraction with the K\"ahler form on $\bbb$ vanishes
\beq
\label{decomp}
F = F^{2,0} + F^{1,1}_- \, .
\eeq
Using the diagonal metric on $\mathbb{T}^4$ associated to the K\"ahler form $J_{\bbb}$, it is 
convenient to define an orthogonal set of two-forms
\bea
\label{dualanti}
&& j_{\pm}^1 = e^1 \wedge e^2 \pm e^3 \wedge e^4 \, ,\nn \\
&& j_{\pm}^2 = e^1 \wedge e^3 \mp e^2 \wedge e^4 \, , \nn\\ 
&& j_{\pm}^3 = e^1 \wedge e^4 \pm e^2 \wedge e^3 \, ,
\eea
such that $j^i_{\pm} = \pm \ast j^i_{\pm}$ (for $i=1,2,3$) and $j^i_{\pm}
\wedge j^j_{\pm} = \pm \frac{1}{2} \delta^{ij} \, \vol (\mathbb{T}^4)$. Then
$J_{\bbb} = j^1_+$ and  $\omega_{\bbb} =  j^2_+ + i j^3_+$.
The decomposition (\ref{decomp}) becomes
\beq
F = f_+ (j^2_+ + i j^3_+) + f_i \, j^i_-
\eeq
for a set of complex $f_+, f_i$. It is not hard to verify now that $f_+ = 1, f_i = 0$ for $n$ 4.7, 
$f_+ = 1, f_i
= (0,1,0)$ for $n$ 4.4, $f_+ = \frac{1}{2}, f_i = (0,\frac{1}{2},\frac{i}{2})$ for $n$ 4.6, and $f_+ = -\frac{1+3i}{2}, f_i = (0,\frac{i-3}{2},\frac{1-3i}{2})$ for $n$ 4.5. Hence the 
curvatures for these three cases satisfy the conditions needed to preserve supersymmetry. 

When $F$ is purely imaginary (real) we get a special case 
of a single non-trivial circle fibration. Indeed after setting to zero $f_+$ and the real part of $f_i$, 
the algebra $n$ 4.6 becomes
(0,0,0,0,0,$\frac{1}{2} \times$ (14-23)), which is isomorphic to $n$ 5.1.  Similarly, either by setting 
to zero $f_+$ and the 
imaginary part of $f_i$  in $n$ 4.6 (modulo the factor $\frac{1}{2}$), or by simply setting to zero $f_+$  
in $n$ 4.4, one gets a nilpotent algebra (0,0,0,0,13+24,0) which is again isomorphic to $n$ 5.1. 
For 4.5 one of the two $U(1)$'s can also be chosen trivial; the non-
trivial fibration will be in a direction that is a linear combination of  $x^5$ and $x^6$.
 We conclude by recalling again that all type C solutions on each of these nilmanifolds can also  be 
obtained by ordinary T-duality from a type B solution with a specific choice of NSNS flux.



\subsubsection{Iterating the twist}

The list of IIB $SU(3)$ structure solutions with O5/D5 sources on nilmanifolds includes only one case which is not related by a 
sequence of T-dualities to  flux compactifications on straight $\mathbb{T}^6$ \cite{GMPT06}. The existence
of such isolated solution is somewhat puzzling, and, as we shall see, it is
related to the rest of nilmanifold compactifications by the twist transformation.\\

The manifolds $n$ 4.3 and 4.6 have trivial $S^1$ factors. These  
can be twisted as well, moving us in the table of
nilmanifolds into the domain of lower $b_1$. In particular 
$n$ 4.6 has the form $M = S^1 \times N_5$ where $N_5$ is a $\mathbb{T}^2$ fibration over 
$\mathbb{T}^3$.  The second cohomology of $N_5$ is non-trivial ($b_2(N_5)=6$) and hence 
it can support non-trivial
$U(1)$ bundles. A priori there can be up to six different ways of 
constructing a $U(1)$ fibration and there are several topologically distinct ways to produce 
a manifold with $b_1(M)$=3 out  of $n$ 4.6. However we will see that
one of them is singled out by supersymmetry.

In the previous section it was shown that $n$ 4.6 yields a type IIB solution 
with O5/D5 sources. We want to further twist the remaining $U(1)$ bundle without
changing the type of sources. This requires  taking a real twist of the  $S^1$ factor while  $B$-transforming with a closed $B$.
From the $n$ 4.6 algebra (0,0,0,0,13,14) it is not hard to see that the 
$S^1$ corresponds to the direction 2, and hence the twisting amounts to sending ${\rm d}\tilde{e}^2 = 0$ to ${\rm d}\tilde{e}^2 = F$ where $F \in
H^2(N_5)$. The algebra becomes (0,$F$,0,0,13,14)\footnote{It is easy to check that $F$ is a linear combination of $e^1\w e^5$, $e^1\w e^6$, 
$e^3\w e^5$, $e^3\w e^4$, $e^4\w e^6$ and $e^3\w e^6 + e^4\w e^5$.}.
The form of the $F$ is again fixed by imposing that the supersymmetry 
equations \eqref{O5cond} continue to hold. This yields the conditions
\bea
F \w (e^3 + ie^4) \w (e^5 + ie^6) &=& 0 \, ,\nonumber \\
F \w( e^1 \w e^3 \w e^4 + e^1 \w e^5 \w e^6) &=& 0 \, ,
\eea
which are solved by $F =  \tilde{e}^3 \w  \tilde{e}^5 +  \tilde{e}^4 \w  \tilde{e}^6$, where we set 
$ \tilde{e}^i = e^{A} e^i$ for $i=1,\ldots,4$ and $ \tilde{e}^i = e^{-A} e^i$ for $i=5,6$. The corresponding 
algebra is (0,35+46,0,0,13,14), which is indeed isomorphic to $n$ 3.14, (0,0,0,12,23,14 - 35). 
In \cite{GMPT06} it was shown that $n$ 3.14 
corresponds to the only solution involving
nilmanifolds that was not obtained by T-duality from compactifications on $\mathbb{T}^6$ with fluxes. 
Our twist transformation does connect it to the rest of the nilmanifold solutions family.\\

A typical feature of such non T-dual solutions is that they involve non-localised intersecting sources, in this case two O5 planes.
It is easy to see that our twist leads to the same result. Indeed, the Bianchi identity for the $F_3$ flux
\beq
\label{BIF3}
g_s {\rm d}F_3= 2 i \del \bar \del (e^{-2A} J) = \delta(D5)-\delta(O5)  \ , 
\eeq
with the K\"ahler form
\beq
J = e^{-2 A} (\tilde{e}^1 \w \tilde{e}^2 + \tilde{e}^3 \w \tilde{e}^4) + e^{2 A} \tilde{e}^5 \w \tilde{e}^6 \, ,
\eeq 
becomes
\bea
g_s {\rm d} F_3 &=& 2 [ \nabla(e^{-2 A}) - e^{2 A}] \tilde{e}^1 \w \tilde{e}^2 \w \tilde{e}^3 \w \tilde{e}^4 +
                       2 e^{-2 A} \, \tilde{e}^3 \w \tilde{e}^4 \w \tilde{e}^5 \w \tilde{e}^6 \nn \\
                && +{\rm d}(e^{-2 A}) (\tilde{e}^2 \w \tilde{e}^4 \w \tilde{e}^6 + \tilde{e}^2 \w \tilde{e}^3 \w \tilde{e}^5) 
                +  {\rm d}(e^{2 A}) (\tilde{e}^2 \w \tilde{e}^4 \w \tilde{e}^5 - \tilde{e}^2 \w \tilde{e}^3 \w \tilde{e}^6)  \, . 
\eea
In order to be consistent with the calibration conditions for the sources, the last line should vanish. 
Had we assumed that $\partial_5$, $\partial_6$ and $\partial_2$ are all honest isometries, this would 
set $A=$const., thus giving the unsurprising result that due to the intersection the sources are smeared.\footnote{Notice
 that while keeping the transformation real ensures that there is no change in the type of solution and hence the sources (both the solution involving $n$ 4.6 and the one on $n$ 3.14 are of type C and have O5/D5 sources), relative orientations of individual sources can change.} There is however the possibility of keeping the $x^2$ dependence in the warp factor and still have a 
consistent (and partially localized) solution. This possibility assumes that the last twist (in direction 2) did not really require an isometry but a circle action. This also suggests a possible generalization of our procedure, but we shall not pursue this further.

\section{Heterotic transformations}\label{het}

In this section we will apply the twist transformation to the heterotic string. Heterotic string 
provides the first examples where compactifications with non trivial
NSNS fluxes have been studied in full detail \cite{S, Hu}. 
We shall consider here the twist transformation on non-trivial flux backgrounds preserving at least 
$\mathcal{N}=1$ supersymmetry. The internal manifold will always be locally a product of $K3$ and 
$\mathbb{T}^2$.  As discussed in \cite{BTY, Se} a chain of dualities can relate a solution involving 
$K3 \times \mathbb{T}^2$ to one where the internal space is given by a non-trivial $\mathbb{T}^2$ 
fibration over $K3$. It is natural to ask whether they could be related
by an $O(6,6)$ transformation of the type \eqref{eq:O(6,6)tr}.

As we discussed in the Section \ref{odd}, the action of  $O(6,6)$  is naturally implemented in the 
Generalized Geometry framework. Such an approach is missing for the heterotic string, basically 
because of the absence of a good twisting of the exterior derivative.
It is nevertheless possible to derive differential equations on pure spinors that capture completely the information contained in the supersymmetry variations. This is all we need
to act with the  $O(6,6)$ transformation \eqref{eq:O(6,6)tr}. In this section we will derive the
equations for the pure spinors in the heterotic string and use them to build the  $O(6,6)$ transformation
connecting the $SU(3)$ structure solutions of \cite{BTY}.

\subsection{$\mathcal{N}=1$ supersymmetry conditions}

Before writing the pure spinor equations for $\mathcal{N}=1$ compactifications in the heterotic case, we
will briefly recall the conditions for $\mathcal{N}=1$ supersymmetry  \cite{S, Hu}.

The supersymmetry equation for the heterotic case can be written\footnote{These conventions are the same as in type II \cite{GMPT04} with the RR fluxes set to zero. Note that these are related to the conventions of
\cite{BBFTY} via $H \rightarrow -H$.}
\bea
&& \delta \psi_M = (D_M -  \frac{1}{4} H_M ) \epsilon = 0 \, ,\nn\\
&& \delta \lambda =  (\not \! {\partial} \phi - \frac{1}{2} \not \! H  ) \epsilon = 0 \, ,\nn\\
&& \delta \chi =  2 \not \! \fff \epsilon = 0 \, ,
\label{hetsusy}
\eea
where $\epsilon$ is a positive chirality ten-dimensional spinor. $\fff$ is the gauge field strength taken to be hermitian\footnote{Following conventions of \cite{BBFTY}, we can develop the gauge quantities in terms of hermitian generators $\lambda^a$ in the vector representation of $SO(32)$, and we use the normalisation condition $tr(\lambda^a\lambda^b)=2\delta^{ab}$.}, i.e. defined with the following covariant derivative on the gauge connection $\aaa$
\beq
\fff=({\rm d} - i \aaa \w)\aaa \ .
\eeq
The conditions that  $\mathcal{N}=1$ supersymmetry imposes on compactifications to a four-dimensional 
maximally symmetric space and non trivial NSNS flux were derived in  \cite{S}. 
If we write the ten-dimensional string frame metric as in type II, \eqref{metricansatz},
\beq
{\rm d}s^2 = e^{2 A} h_{\mu\nu} {\rm d}x^\mu {\rm d}x^\nu +g_{m n}(y) {\rm d}y^m {\rm d}y^n \, , 
\eeq
then the warp factor must be zero $A=0$ and the four-dimensional metric Minkowski
\beq
h_{\mu \nu} = \eta_{\mu \nu} \, .
\eeq 
The internal manifold must be complex.
The holomorphic three-form $\Omega$ satisfies
\beq
{\rm d} (e^{- 2 \phi} \Omega ) = 0 \, . \label{susyomega}
\eeq
In terms of the complex structure $I$ defined by $\Omega$, the K\"ahler form is $J_{mn} = I^{\ \ p}_m g_{pn} $ and satisfies
\bea
&& {\rm d}J = i (H^{1,2} - H^{2,1} ) \Leftrightarrow H=i(\partial - \ov{\partial})J \, , \label{dcj}\\
&& {\rm d} (e^{-2 \phi} J \wedge J ) =0 \, . \label{confclosed}
\eea
The NSNS three-form has only components $(2,1)$ and $(1,2)$ with respect to the complex structure $I^{\ \ m}_n$
\beq
H = H^{2,1}_0 + H^{1,2}_0 + ( H^{1,0} + H^{0,1}) \wedge J \, , \label{Hdec}
\eeq
where the subindex $0$ denotes the primitive part of $H$.

The gauge field strength $\fff$ must satisfy the six-dimensional hermitian Yang-Mills equation, i.e. must be of type $(1,1)$ and primitive
\bea
&& \fff \llcorner J = 0 \, ,\\
&& \fff_{i j} = \fff_{\bar{i} \bar{j}} = 0 \, ,
\eea
where the second equation is given in holomorphic and anti-holomorphic indices.\\

These are the necessary conditions imposed by supersymmetry. The equations of motion are satisfied provided  the Bianchi identity holds:
\beq
H={\rm d}B -\frac{\alpha^\prime}{4} tr\left(\aaa\w {\rm d}\aaa - i\frac{2}{3} \aaa \w \aaa \w \aaa \right)+ \frac{\alpha^\prime}{4} \omega_3(M) \ , \label{db}
\eeq
where $\aaa$ is the gauge connection and $\omega_3(M)$ the Lorentz Chern-Simons term \cite{CHSW}.  It is easier to check the anomaly cancellation condition
\beq
{\rm d}H=2i\overline{\partial}\partial J=\frac{\alpha^\prime}{4}[tr(\mathcal{R}\w \mathcal{R}) - tr(\fff \w \fff)] \ .
\eeq

\subsection{Pure spinor equations for heterotic compactifications}

In the four plus six-dimensional splitting, the supersymmetry parameter $\epsilon$ corresponds, 
for $\mathcal{N}=1$ supersymmetry, to a single six-dimensional chiral spinor $\eta_+$
\beq
\label{hetdecompo}
\epsilon =  \zeta_+ \otimes \eta_+ + \zeta_- \otimes \eta_- \, ,
\eeq
where $\zeta_+$ is, as always, a four-dimensional Weyl spinor of positive chirality 
($\zeta_- = (\zeta_+)^\ast$) and  $\eta_- = (\eta_+)^{\ast} $. 
The spinor $\eta_+$ can be seen as defining an $SU(3)$ structure on $M$ (and indeed the supersymmetry 
conditions can be rephrased in terms of conditions on the torsion classes of an $SU(3)$ structure 
manifold). Then a natural choice for the pure spinors is
\bea
\label{Phiphet}
\Psi_+ &=&  8 \, e^{-\phi} \eta_+ \otimes \eta^{\dagger}_+ =  e^{-\phi} \, e^{-iJ} \, , \nn \\
\Psi_- &=& 8 \, e^{-\phi}\eta_+ \otimes \eta^{\dagger}_-  = -i e^{-\phi} \Omega \, .
\eea
We have used the same letter as in \eqref{eq:Psiplus} for the fermion bilinears (\ref{Phiphet}), and we will still 
call them  pure spinors. However it should be kept in mind 
that they are not defined on the generalized tangent bundle $E$ but on $T \oplus T^*$ 
($e^{- B}$ is missing). Using (\ref{hetsusy}) and (\ref{hetdecompo}), one can obtain the 
supersymmetry conditions on the pure spinors \cite{GMPT04, GMPT06}
\beq 
{\rm d}\left( \Psi_{\pm} \right)= H \bullet \Psi_{\pm} \ , \label{SUSYdot}
\eeq
with
\beq
H \bullet \Psi_\pm = \frac{1}{4}  H_{mnp} ({\rm d}x^m \w {\rm d}x^n \w i^p  
- \frac{1}{3} i^m i^n i^p ) \Psi_\pm \ .
\eeq

Even though (\ref{SUSYdot}) captures all the information contained in supersymmetry variations, there 
are two problems with the action of the $({\rm d}-H \bullet)$ operator: it is not a differential, and 
it is hard to interpret its action on pure spinors as a twisting. There is a partial resolution to the 
former problem. The  $\Psi_-$ equation yields that $H$ is indeed only of  (1,2) + (2,1) type as given 
in (\ref{Hdec}), and
\beq \label{cs-int}
{\rm d}\left( \Psi_- \right)=   i H^{0,1} \wedge \Psi_- \, ,
\eeq
from which we conclude that the internal manifold is complex.  We can now use the integrability of the complex structure (\ref{cs-int}) to rewrite (\ref{SUSYdot}) in terms of a differential
\beq
\label{ps}
{\rm d}\left( \Psi_{\pm} \right)=  \pm \left[ (H^{1,2} - H^{2,1}) - i (H^{0,1}-H^{1,0})\right] \wedge \Psi_{\pm} = \pm {\check H} \wedge \Psi_{\pm} \, .
\eeq

The equation (\ref{ps}) for  $\Psi_-$  agrees with (\ref{cs-int}). The decomposition 
of the   $\Psi_+$ equation by the rank of the differential forms  gives
\begin{itemize}
 \item at degree 1
\beq \label{rank1}
{\rm d} \phi = i (H^{0,1}-H^{1,0}) \, ,
\eeq
using which we recover the correct scaling on $\Omega$ (\ref{susyomega}).
 \item at degree 3
\bea
 {\rm d} J &=& i (H^{1,2} - H^{2,1}) \label{rank3}\\
 &=& i ( H_0^{1,2} - H_0^{2,1} ) + {\rm d} \phi \wedge J \, . \label{r3}
\eea
Eq.  (\ref{rank3}) is clearly (\ref{dcj}). Wedging (\ref{r3}) with $J$, we recover the balanced metric 
condition (\ref{confclosed}).  Finally recalling that 
$*H = i(H_0^{2,1} - H_0^{1,2} - H^{1,0}\wedge J + H^{0,1} \wedge J)$ we arrive at
\beq
\ast H = - e^{2\phi} {\rm d}(e^{-2\phi} J) \ .
\eeq
 \item at degree 5, there is no new information.
\end{itemize}

\noindent
We can now check that ${\rm d}\mp {\check H} \w$ is a differential. Since $\check H$ is made of odd 
forms,  it squares to zero, and, due to (\ref{rank1}) and (\ref{rank3}), ${\rm d}{\check H}=0$. Hence  
$({\rm d} \mp {\check H} \w)^2=0$. \\

There stays however the problem that we cannot see the action of ${\rm d}\mp {\check H} \w$ as a 
result of a twisting on the pure spinor. This will not prevent us for using the twist 
transformation to relate 
different heterotic backgrounds. Essentially the idea is to consider a very special case of the 
transformation (\ref{gentwist}) which does not contain a $B$-transform nor changes 
the phase of the pure 
spinor (even if this amounts to  stepping back somewhat from the Generalized Geometry). 
In other words, 
we keep  only the twist part of the general transformation (\ref{gentwist}) and we demand that
\beq
({\rm d}\mp {\check H}^\prime \w) (O_c \, \Psi_{\pm}) =0 \, .
\eeq
Two internal geometries $M$ and $M'$, defined by the pairs $ \Psi_{\pm}$ and 
$\Psi'_{\pm} = O_c \, \Psi_{\pm}$, are related via twisting and satisfy the same type of 
${\check H}$-twisted integrability conditions. The pair of manifolds connected this way may in general 
be topologically and geometrically distinct. Examples of such connections were constructed recently 
in \cite{swann}. Since there is no $B$-transform involved in the construction, we are not dealing
 here with the diffeomorphisms of the  generalized tangent bundle. In this sense the discussion of 
the heterotic string differs from the rest of the paper.

\subsection{SU(3) structure solutions}\label{solhet}

We shall return to the class of fibered  metrics discussed earlier. Consider a six-dimensional internal 
space with a four-dimensional base $\bbb$ which is a conformal Calabi-Yau, and a $\mathbb{T}^2$ fibre 
with holomorphic coordinate $z=\theta^1 + i\theta^2$. The metric and the $SU(3)$ structure on the 
internal space are in general given by
\bea
{\rm d}s^2&=&e^{2\phi} {\rm d}s_{\mathcal{B}}^2+ \Theta \overline{\Theta}  \, ,   \nn\\
J&=&e^{2\phi} \, J_{\mathcal{B}} + \frac{i}{2} \Theta\w \overline{\Theta} \label{J}  \nn \\
\Omega&=&e^{2\phi} \, \omega_{\mathcal{B}}\w \Theta \label{Omega}
\eea
where $\Theta={\rm d}z+\alpha$ and $\alpha$ is a $(1,0)$ connection one-form. 
$J_{\mathcal{B}}$ is the CY K\"ahler form, $\omega_{\mathcal{B}}$ is the CY holomorphic two-form, 
and the dilaton $\phi$ depends only on the base coordinates. Furthermore, the curvature of 
the $\mathbb{T}^2$ bundle $F={\rm d}\alpha$ has to be primitive with respect to $J_{\mathcal{B}}$
\beq \label{primi}
F \w J_{\bbb}=0 \, , \qquad \mbox{and} \qquad F \w \omega_{\bbb}=0 \, .
\eeq
A general solution to these constraints is of the form 
$F = F^+_{(2,0)} + F^-_{(1,1)} \in H^{2,+}(\bbb) \oplus H^{2,-}(\bbb)$. Then, one can satisfy the 
local supersymmetry equations, provided the base $\bbb$ is a four-dimensional hyper-K\"ahler surface. Here,
the equations (\ref{susyomega}) and (\ref{confclosed}) are automatically satisfied. \\

In \cite{BTY}, two $\mathcal{N}=2$ solutions with $\bbb=K3$ and a non-zero $H$ have been discussed.
In the first solution (which we will denote by Solution 1),
the internal manifold is the direct product $K3 \times \mathbb{T}^2$ , i.e. $\alpha=0$. 
The gauge bundle is reduced to the sum of $U(1)$ bundles, so $\fff$ is a sum of (1,1) primitive 
two-forms on the base. Furthermore, in this solution, $B=0$, so $H$ receives only $\alpha'$ 
contributions. The dilaton is non-trivial and the condition (\ref{dcj}) relates its derivatives 
to the gauge term.

The second solution (Solution 2) consists of a non-trivial $\mathbb{T}^2$ fibration\footnote{The Betti 
numbers 
are $b_1(M)=0$, $b_2(M)=20$ and $b_3(M) = 42$. Note that the Euler number $\chi(M)$ vanishes, thus
the manifold has a global $SU(2)$ structure.} over $K3$, so we have an $\alpha \neq0$. 
Moreover  $\fff=0$ and $B=\re(\alpha\w {\rm d}\ov{z})\neq 0$. The 
dilaton is non-trivial, and has the same value as in the previous solution. The curvature of the 
connection $\alpha$ is in general given by \eqref{decomp}, and the solution would then be 
$\mathcal{N}=1$. 
If $F$ has only a (1,1) part as in \cite{BTY}, the solution is $\mathcal{N}=2$.\footnote{The $\mathcal{N}=2$ supersymmetry 
is easy to see using the $SU(2)$ structure. There exists a second pair of compatible pure spinors 
which are of type 1-2, namely $\Psi_+ = e^{-\phi} \omega_{\bbb} \w \exp(\Theta \w \ov{\Theta}/2)$ and 
$\Psi_- = e^{-\phi} \Theta \w \exp(- iJ_{\bbb})$ (where we chose $\theta_+=\frac{\pi}{2}\ , \ \theta_-=\pi$). Differently from the type 0-3 pair, now it is 
$\Psi_-$ which is not closed. The closure of $\Psi_+$ imposes a stronger condition than 
(\ref{primi}) requiring that $\omega_{\bbb} \w F^I = 0$ (for $I=1,2$) hence restricting 
$F = F^-_{(1,1)} \in H^{2,-}(\bbb)$.} \\

These two solutions were proven to be related by a  transition  \cite{DRS, BD, BBFTY, BTY, Se}. 
Both solutions arise from M-theory compactifications on $K3\times K3$. 
A first step consists in reducing to type IIB solutions on an orientifold 
$(\mathbb{T}^4/\mathbb{Z}_2) \times (\mathbb{T}^2/\mathbb{Z}_2)$. This is achieved by taking 
the two $K3$ at the point in moduli space where they both are 
$\mathbb{T}^4/\mathbb{Z}_2$ orbifold. 
Then one of the two $\mathbb{T}^4/\mathbb{Z}_2$ is considered as a fibration of $\mathbb{T}^2$ over
$\mathbb{T}^2/\mathbb{Z}_2$, and the area of the fibre is taken to zero. This yields  
a type IIB solution on $(\mathbb{T}^4/\mathbb{Z}_2) \times (\mathbb{T}^2/\mathbb{Z}_2)$ with four 
$D7$ and one $O7$ at each of the four fixed points of $\mathbb{T}^2/\mathbb{Z}_2$. 
Then two T-dualities along $\mathbb{T}^2/\mathbb{Z}_2$ give
a dual type IIB solution on $(\mathbb{T}^4/\mathbb{Z}_2) \times (\mathbb{T}^2/\mathbb{Z}_2)$ 
with $D9$ and $O9$ at the dual points. The same solution can also be interpreted as a type I 
solution on $K3 \times \mathbb{T}^2$ where $K3$ is understood as 
$\mathbb{T}^4/\mathbb{Z}_2$. Finally, doing an S-duality, one gets the heterotic $SO(32)$ solution on $K3 \times \mathbb{T}^2$ where $K3$ is again understood as $\mathbb{T}^4/\mathbb{Z}_2$. 
The transition between the two heterotic solutions then corresponds to an exchange of the two $K3$, and of its $(1,1)$ two-forms, namely 
$\fff$ and $F$. Note that M-theory on 
$K3 \times K3$ can be dual to type IIA on $X_3 \times S^1$ where $X_3$ is a CY three-fold. Then, the 
exchange of the two $K3$ corresponds to mirror symmetry for $X_3$ \cite{A}. This exchange should 
not change the dilaton, which is therefore the same in the two solutions.\\

We may connect these two solutions directly via (the special case of) our transformation 
\eqref{OtrspT2}. Since we have a background with only two commuting isometries, the twist takes the form
\beq
\label{ohet}
O_c=1+o=1+\alpha \w i_{\partial z} + \overline{\alpha} \w i_{\partial \overline{z}}+\alpha \w \overline{\alpha} \w i_{\partial \overline{z}}\ i_{\partial z} \, .
\eeq
It has the effect of sending $dz$ to $\Theta ={\rm d}z + \alpha$ in the forms defining the $SU(3)$ structure 
(\ref{Omega}), and hence it relates the internal geometries of two solutions. 
Since the only change in the metric between the two solutions is the presence of a non-trivial connection,
we did not assume any rescaling of the metric and thus we set 
$A_\bbb=\mathbb{I}_4$ and $A_\fff= \mathbb{I}_2$ in \eqref{OtrspT2}. As a consequence the dilaton does not
change, in agreement with the analysis of  \cite{BTY}.

Thus, starting with Solution 1 we read off the $H$ from the closure of the transformed pure spinor
\bea
H&=&i(\partial - \ov{\partial})J \ = i(\partial - \ov{\partial})(e^{2\phi})\w \ J_{\mathcal{B}} - \frac{1}{2} (\partial - \ov{\partial}) \left( ({\rm d}z+\alpha)\w ({\rm d}\ov{z}+\ov{\alpha}) \right) \nn\\
&=&i(\partial - \ov{\partial})(e^{2\phi})\w \ J_{\mathcal{B}} - \frac{1}{2} (\partial - \ov{\partial}) \left( \alpha\w \ov{\alpha} \right)+ {\rm d} \left( \re(\alpha\w {\rm d}\ov{z})\right) \, ,
\eea
where we used the anti-holomorphicity of $\alpha$. The last term is the only closed part of $H$, 
and comparing to \eqref{db} we derive the $B$-field of Solution 2
\beq
B=\re(\alpha\w {\rm d}\ov{z}) \ .
\eeq
Furthermore,
\beq \label{dh-het}
{\rm d}H =  -2i \partial \ov{\partial} (e^{2\phi}) \w J_{\mathcal{B}} + F \w \ov{F} \, .
\eeq

We would like to stress once more that the two solutions were related using the 
transformation on the tensor products  (\ref{Phiphet}).  Differently from the pure spinors in type II 
solutions these do not contain an $e^{-B}$ factor and we have not performed any $B$-transform 
in mapping the solutions; rather the $B$-field was read off as the closed part of $H$. \\

The global aspects of the solutions deserve some comments. 
Eq.\eqref{dh-het} has the same 
structure as the tadpole condition for the O5/D5 solutions in type IIB. Notice that, in general, the 
first term in  (\ref{dh-het}) yields $\delta$-function contributions which are associated with the 
positions of branes and planes, while the second term, after being completed to a top-form by wedging with $J$, integrates over the six-manifold $M$ to a positive number. 
The presence of these defects is what makes $\mathbb{T}^2$ fibrations over $\bbb=\mathbb{T}^4$ an 
admissible basis for the solutions in IIB. In heterotic string in the absence of good candidates for 
negative tension defects, we would like to assume a smooth dilaton; the second term is then cancelled 
by the $\alpha'$ contributions to (\ref{db}).  
Crucially, when $\bbb = K3$, terms like 
$\int_M \partial \ov{\partial} (e^{2\phi}) \w J^2$   vanish for any smooth $\phi$, while for 
$\bbb = \mathbb{T}^4$,  $\phi$ may be non-single valued and the integral gives a finite contribution 
to the tadpole. Indeed it is known that compactifications on  smooth $\mathbb{T}^2$ fibrations over $\mathbb{T}^4$ are forbidden by the heterotic Bianchi identity \cite{FuY, BBFTY}. Starting from a heterotic compactification on $\mathbb{T}^6$ and applying 
the transformation (\ref{eq:O(6,6)tr}) with non-single valued coefficients (and hence the dilaton) may allow to circumvent the constraints imposed by the Bianchi identity. However such backgrounds will be non-geometric and we will not discuss them further in this paper.\\

We conclude this section by turning briefly to the transformation of the gauge field $\fff$. The ordinary $O(2,18)$ transformation on the Narain lattice can exchange the antiself-dual part of the 
curvature of the $\mathbb{T}^2$ fibration with the $U(1)$ factors in the gauge bundle. This exchange is 
consistent both with supersymmetry and tadpole cancellation.
As discussed in \cite{EM}, a better understanding of this exchange, as well as the transformation of the 
$\alpha^\prime$ terms of $H$, is achieved considering the pullback of $H$ to the total space of 
the gauge bundle $\rho$: $\mathcal{P} \rightarrow M$, 
$$\mathcal{H} = \rho^* H - \frac{\alpha'}{4} tr \aaa \w \fff \, ,$$ whose contraction with the isometry 
vectors $\partial_z$ and $\partial_{\ov{z}}$ gives a closed two-form (which can be exchanged with the 
gauge $U(1)$ curvature terms). 
For our purposes, in order to capture the transformation of the $\alpha'$ terms,
one possibility is to extend the $O(d,d)$ group to $O(d+16,d+16)$ transformations, and introduce new 
generalized vielbein incorporating the gauge connection. We discuss this possibility in 
Appendix \ref{Fhet}.

\section{Courant bracket and a coordinate dependent $O(n,n)$ transformation}\label{Courant}

In this section we shall provide, from a different point of view, some additional, a posteriori justification for the transformations argued in this paper. Inspired by the twist, we shall consider a theory with a $\mathbb{T}^n$ action and with an integrable generalized complex structure and discuss the possibility of existence of coordinate dependent $O(n,n)$ transformations, that may preserve the integrability of the generalized complex structure.

It is well known that global $O(n,n,\mathbb{Z})$ is a symmetry of equations of motion. 
Indeed, a multiplication by constant  $O(n,n,\mathbb{Z})$ matrices exchanges sigma model 
equations of motion with Bianchi identities while leaving the whole system invariant. This
 symmetry should be an automorphism of the sigma model current algebra. Given a section 
$(v,\rho)$ of $TM\oplus T^*M$ one can construct a current
\begin{equation}
J_{\epsilon}{(v,\rho)}=\oint_{S^1}{\rm d}\sigma\,\epsilon(\sigma)
\bigl[\imath_vp+\imath_{\partial_{\sigma}x}\,\rho\bigr]
\end{equation}
where $x$ are coordinates on $M$, $p$ are momenta and $\epsilon(\sigma)$ is a smooth
 (test) function on the circle (see \cite{Al} for details). The Poisson bracket of two 
such currents is
\begin{equation}
\{J_{\epsilon_1}(v,\rho),\,J_{\epsilon_2}(w,\lambda)\}
=J_{\epsilon_1\epsilon_2}\bigl([(v,\rho),(w,\lambda)]_H\bigr)
-\frac12\oint_{S^1}{\rm d}\sigma\,(\epsilon_1\partial_{\sigma}\epsilon_2-\epsilon_2\partial_{\sigma}\epsilon_1)
\bigl[\imath_{v}\lambda+\imath_{w}\rho\bigr] \, .
\label{JJ}
\end{equation}
$[\cdot,\cdot]_H$ is the twisted Courant bracket and it is defined by
\begin{equation}
[(v,\rho),(w,\lambda)]_H=[v,w]
+\Bigl\{\mathcal{L}_{v}\lambda-\mathcal{L}_{w}\rho
-\frac12\,{\rm d}(\imath_{v}\lambda-\imath_{w}\rho)
+\imath_{v}\imath_{w} H
\Bigr\} \, ,
\end{equation}
where the first term is the Lie bracket of two vectors fields. The only automorphisms of the Courant 
bracket are the diffeomorphisms and closed $B$ transforms. \\

Taking  $M$ to be  a principal torus bundle ($ \mathbb{T}^n\hookrightarrow M \stackrel{\pi} \longrightarrow {\bbb}$), we can study the reduction
of the twisted current algebra to the base $\mathcal{B}$. We start by decomposing the sections of  
$TM\oplus T^*M$ into horizontal and vertical components.  Any vector $v$ and one-form $\rho$ can be written as
\begin{align}
v &= v _{\mathcal{B}}+  f^I K_I   \nonumber \\
\rho &= \rho_{\mathcal{B}}  + \phi_I \Theta^I \, ,\nonumber
\end{align}
where as before, we denote the connections on $\mathbb{T}^n$ as $\Theta^{I=1,...,n}$ and their curvatures as $F^I$ ($\pi^* F^I={\rm d} \Theta^I$). The torus generators are denoted $K_I$ ($\iota_K \Theta=1$). Demanding that both ${\cal L}_K v_{\mathcal{B}}= 0$ and  
${\cal L}_K \rho_{\mathcal{B}} = 0$, implies in particular  $f \in \Omega^0({\mathcal{B}},\mathfrak{t})$ 
and $\phi \in \Omega^0({\mathcal{B}},  \mathfrak{t}^*)$.
In other words, a $\mathbb{T}^n$-invariant section of $TM$ can be written as an element 
$(v_{\mathcal{B}}, f) \in
T\mathcal{B} \oplus \mathfrak{t}$, while a  $\mathbb{T}^n$-invariant section of $T^*M$ can be written 
as $(\rho_{\mathcal{B}}, \phi) \in T^*\mathcal{B}\oplus  \mathfrak{t}^*$.  From now on, we shall drop 
the subscript ${\mathcal{B}}$ on the vectors and one-forms; these will be taken to be horizontal.

Here we  shall be interested only in configurations where the $B$-field has no components with
two legs on the fibre
\beq
B= B_2 + (B_1)_I \wedge \Theta^I \, , \qquad H = \pi^* H_3 + (\pi^* H_2)_I \Theta^I \, .
\eeq
The reduction of the Courant bracket is then pretty simple and is given by
\begin{align}
\label{red-courant}
[(v,f ;  & \rho, \phi),   \,(w, g;   \lambda, \omega)]_{(H_3,F,H_2)}=
[(v;  \rho ),  \,(w; \lambda)]_{H_3} + \nonumber \\
& \Bigl(0,{\cal L}_{v} g - {\cal L}_{w} f ; \langle \omega, {\rm d} f\rangle -  
\langle  \phi, {\rm d} g \rangle  - \frac{1}{2}{\rm d}( \langle \omega, f \rangle  -  \langle  \phi, g \rangle  ), {\cal L}_{v} \omega - {\cal L}_{w} \phi   \Bigr) +  \\
&\Bigl( 0,  \imath_{v}  \imath_{w} F  ;   \langle  \omega ,  \imath_{v} F \rangle + \langle  \imath_{v} H_2 , g\rangle  -  \langle   \imath_{w} H_2 , f\rangle  - \langle  \phi , \imath_{w} F  \rangle ,   \imath_{v}  \imath_{w}  H_2  \Bigr)   \, ,\nonumber
\end{align}
where  $ \langle \cdot, \cdot \rangle$ denotes the natural pairing $\mathfrak{t}^*\otimes\mathfrak{t}\to \mathbb{R}$: $ \langle \omega, f \rangle = \omega_I f^I$ and so on.  For details and derivations 
see \cite{BHM}. \\

It is not hard to see that the one-form part of this expression is invariant under
\beq
\label{auto-dd}
  \left (\begin{array}{c}  f^I \\ \phi _I   \end{array} \right )  \rightarrow   \left (\begin{array}{c c}  A^I{}_J & B^{IJ} \\ C_{IJ} & D_I{}^J \end{array} \right ) \cdot \left (\begin{array}{c}  f^J \\  \phi_J   \end{array} \right ) \qquad \mbox{and} \qquad    \left (\begin{array}{c}  F^I \\ H _I   \end{array} \right )  \rightarrow   \left (\begin{array}{c c}  A^I{}_J & B^{IJ} \\ C_{IJ} & D_I{}^J \end{array} \right ) \cdot \left (\begin{array}{c}  F^J \\  H_J   \end{array} \right )\, ,
  \eeq
provided  $A,B,C,D$ satisfy the relations needed to make the transformation an element of 
$O(n,n,\mathbb{Z})$. Indeed
$ \langle  \omega ,  \imath_{v} F \rangle + \langle  \imath_{v} H_2 , g\rangle$ and   
$ \langle   \imath_{w} H_2 , f\rangle  + \langle  \phi , \imath_{w} F  \rangle $ are separately 
invariant, while the third entry of the second line of (\ref{red-courant}) can be rewritten as 
$\langle \omega, {\rm d} f\rangle  +   \langle   {\rm d}\phi,  g \rangle  
- \frac{1}{2}{\rm d}( \langle \omega, f \rangle  +  \langle  \phi, g \rangle  )$ and its invariance 
is checked readily. The action of  (\ref{auto-dd}) on the second and fourth entries of 
the quartet shows that a constant $O(n,n,\mathbb{Z})$ transformation is indeed an automorphism of the reduced current algebra (\ref{red-courant}).\\

In order to generalize this action to a local $O(n,n)$ transformation it is useful to examine  
(\ref{red-courant}) closer.  The twisted algebra on global sections of 
$T\mathcal{B} \oplus \mathfrak{t} \oplus T^*\mathcal{B}\oplus  \mathfrak{t}^*$  can be viewed as algebra 
on local sections of the generalized tangent bundle of $M$. Let us start from the untwisted 
Courant bracket
\begin{align}
\label{red-courant-simple}
[(v,f ;  & \rho, \phi),   \,(w, g;   \lambda, \omega)] =
[(v;  \rho ),  \,(w; \lambda)] + \nonumber \\
& \Bigl(0,{\cal L}_{v} g - {\cal L}_{w} f ; \langle \omega, {\rm d} f\rangle -  \langle  \phi, {\rm d} g \rangle  - \frac{1}{2} {\rm d} ( \langle \omega, f \rangle  -  \langle  \phi, g \rangle  ), {\cal L}_{v} \omega - {\cal L}_{w} \phi   \Bigr) \, .
\end{align}
In addition to the diffeomorphisms, this bracket has two automorphisms:

\begin{itemize}

\item[1.]
Constant $O(n,n)$ transformations on $ \mathfrak{t} \oplus \mathfrak{t}^*$:
\beq
\label{odd-tans}
S_{\mbox{t}}( X )  =  \left (\begin{array}{cc|cc} \mathbb{I} & 0 & 0 & 0  \\ 0 &A& 0 & B \\ \hline 0  & 0& \mathbb{I}& 0 \\ 0&C & 0 & D \end{array}\right )   \left (\begin{array}{c}  v \\ f \\ \hline  \rho \\ \phi \end{array}\right ) \, .
\eeq
Indeed, it is not hard to check that
\beq
\label{symmetry-cour}
[S_{\mbox{t}} (X), S_{\mbox{t}}(Y)] = S_{\mbox{t}} ([X,Y] ) \, .
\eeq

\item[2.]
 Generalized $B$-transforms
\beq
\label{genbtrans}
X \mapsto e^{\hat{B}} X = \Bigl( v , f+ \imath_{v} U; \rho + \imath_{v} b^{\mathcal{B}} + \langle b, f \rangle + \langle \phi, U \rangle, \phi + \imath_{v} b \Bigr)^T
\eeq
with closed  two-form $b^{\mathcal{B}}$ and one-forms  $b$ and $U$.
\end{itemize}

More explicitly this can be seen as a section of the generalized tangent bundle
\beq
\label{sect}
\mathcal{X} =   \begin{pmatrix} \rho_{\alpha},  \phi_I | v^{\alpha}, f^I   \end{pmatrix}   
\left( \begin{array}{cc|cc} e^{\alpha}{}_{\mu} & 0 & 0 & 0  
\\ U^I{}_{\mu} &  \delta^I{}_J & 0 & 0 \\ 
\hline (b^{\mathcal{B}})_{{\alpha} \mu}  & b_{{\alpha}I} & \hat{e}_{{\alpha}}{}^{\mu} & - U_{{\alpha}}{}^J 
\\ b_{I \mu} &0 & 0 & \delta_I{}^J\end{array}\right ) 
\left (\begin{array}{c} {\rm d} x^{\mu} \\ {\rm d} \theta^J \\ \hline \partial_{\mu} \\ 
\partial_J \end{array}\right )= {X}^T \, \eta \,    \mathcal{E} 
\begin{pmatrix} {\rm d} x^{\mu} \\ {\rm d} \theta^J \\ \hline \partial_{\mu} \\ \partial_J 
\end{pmatrix} \, ,
\eeq
where $x^{\mu}$ and $\theta^I$ are the coordinates on the base manifold $\bbb$  and the 
torus fibre respectively. Note that we consider here a special case of (\ref{eq:genvfiber0}) where the metric on $\mathbb{T}^n$ is chosen to be diagonal, and set the $B$-field fibre component to zero.

When $b^{\mathcal{B}}$,  $b$ and $U$ are not flat, the Courant bracket of two such (local) sections of 
the generalized tangent bundle will yield the twisted bracket $[X,Y]_{(H_3,F,H_2)}$
(\ref{red-courant}). Note that the two-forms appearing in the  bracket
(\ref{red-courant}) are the field strengths of local quantities $U^I{}_{\mu} $ and $ b_{\mu I}$,
$F_2 = {\rm d} U$ and $H_2 = {\rm d}b$.

We can now check that the transformation $S_{\mbox{t}} (e^{\hat B} X)$ corresponding to (\ref{auto-dd}) is an automorphism of (\ref{red-courant}):
\beq
\label{symmetry-courplus}
[S_{\mbox{t}} (e^{\hat B}X), S_{\mbox{t}}(e^{\hat B}Y)] = S_{\mbox{t}} (e^{\hat B} [X,Y] ) \, .
\eeq \\

Having described this way the constant $O(n,n)$ symmetry of the current algebra, we are ready to extend 
it to the action of (base) coordinate dependent $O(n,n)$. Consider now $X \mapsto e^{\hat B} S_{\mbox{t}} (X)$ and let us take in general the $O(n,n)$ matrix to be coordinate dependent.
Starting with a manifold $M$ which is a principal torus bundle with a connection 
$\Theta = {\rm d} \theta + U^{(0)}$ and $B$-field $b^{(0)}$, let us choose the components of $e^{\hat B}$ such that  $U = A U^{(0)} + B b^{(0)} + V$ and $b = C U^{(0)} + D b^{(0)} + c$, or in other words
\beq
\label{b-hattrans}
 e^{\hat B} S_{\mbox{t}} (X) =  S_{\mbox{t}} e^{\hat B^{(0)}} (X) + \Bigl( 0, \imath_{v} V;   \langle c, A^T f  + B^T \phi\rangle + \langle C^T f + D^T \phi, V \rangle , \imath_{v} c \Bigr)\, .
\eeq

The conditions for
\beq
\label{symmetry-courplusplus}
[e^{\hat B} S_{\mbox{t}} (X), e^{\hat B} S_{\mbox{t}}(Y)] = e^{\hat B} S_{\mbox{t}} ([X,Y] )
\eeq
now are:
\bea
\label{auto-scal}
 (g^I \imath_v -f^I \imath_w) {\rm d} A^J{}_I   + (\omega_I \imath_v - \phi_I \imath_w)  {\rm d}B^{JI}  +  \imath_v \imath_w \Bigl( {\rm d} V^J  + {\rm d} A^J{}_I U^{(0) I}  + {\rm d} B^{JI} b^{(0)}{}_I \Bigr)&=&  0 \nonumber \\
(g^I \imath_v -f^I \imath_w){\rm d}  C_{JI}   +   (\omega_I \imath_v - \phi_I \imath_w){\rm d}  D_J{}^I  +  \imath_v \imath_w \Bigl( {\rm d} c_J  + {\rm d} C_{JI} U^{(0) I}  + {\rm d} D_J{}^I  b^{(0)}{}_I \Bigr) &=& 0
\eea
In other words, the coordinate dependence of the $O(n,n)$ parameters is compensated by the failure of the algebraic conditions $U - A U^{(0)} - B b^{(0)} =0 $ and $b - C U^{(0)} - D b^{(0)} =0$.

The one-form part yields:
\beq
\label{auto-form}
\begin{pmatrix} g & \omega \end{pmatrix} O^T \eta \, {\rm d}O
\begin{pmatrix} f \\ \phi\end{pmatrix}
+
\Bigl( \begin{pmatrix} g & \omega \end{pmatrix} O^T \eta \, \imath_v
-
\begin{pmatrix} f & \phi \end{pmatrix} O^T \eta \, \imath_w \Bigr)
 \left( {\rm d} O
\begin{pmatrix} U^{(0)} \\ b^{(0)}  \end{pmatrix} +
\begin{pmatrix} {\rm d}V \\ {\rm d}c \end{pmatrix} \right)=0
\eeq

Hence combining the $\hat B$ transform and the $O(n,n)$ rotation of $\mathfrak{t} \oplus  \mathfrak{t}^*$, 
there is a possibility of gaining a new symmetry - a coordinate-dependent automorphism  of the current 
algebra. This is subject to solving (\ref{auto-scal}) and (\ref{auto-form}).  These give conditions on the 
curvatures ${\rm d}V$ and ${\rm d}c$ in terms of the original geometric data. In order to be symmetry of the theory,  \eqref{auto-scal} and \eqref{auto-form} should a priori be 
satisfied for arbitrary sections of $T\bbb \oplus T^*\bbb$ and $\mathfrak{t} \oplus  \mathfrak{t}^*$. We believe that in general the constant $O(n,n)$ and closed generalized $B$-transform, discussed above, are the only solutions.

Notice however that in studying supersymmetric compactifications we are interested in specific 
situations where the manifolds admits at least one integrable generalized complex structure. 
In this case all we need to impose is the closure of the Courant bracket, and
the relative automorphisms, only on the eigenspaces of such generalized complex structure and not
on a generic section of $E$.   Moreover, as discussed in Section \ref{tpII}, 
even if weaker than the closure of the Courant bracket, a sufficient condition for having an integrable 
Generalized Complex Structure is a twisted closure of the corresponding pure spinor. 
The transformation of such a pure spinor under the coordinate-dependent $O(n,n)$ transformation 
and the conditions of its closure are easier to analyse and are of better practical use. 
The analysis of Section \ref{tpII} and the examples of non-trivial supersymmetric solutions are 
special cases for what appears to be a larger automorphism of the Courant bracket.\\

We conclude by remarking once more that the coordinate-dependent $O(n,n)$  symmetry is ``bigger'' than 
constant $O(n,n, \mathbb{Z})$, as it might lead to a topology change in situations where the constant 
$O(n,n)$ does not. This is the case for our principal examples, namely the duality transformation from 
$M = \mathcal{B} \times \mathbb{T}^n$ 
with $B^{(0)}=0$ to non-trivially fibered dual geometry $\mathbb{T}^n\hookrightarrow M^\prime \stackrel{\pi}{\longrightarrow} \mathcal{B}$ and ${\tilde B} \neq 0$ for the cases where the torus fibre has
dimensions $n=2$ and $n=1$.
Moreover our examples have been based on the simplest choices of solutions (for example 
the operator $O$ in \eqref{OtrspT2} acts diagonally on the fibre). It should be interesting to obtain the  general conditions for the coordinate dependent twist-like automorphisms of the current algebra and consider more intricate examples of dual backgrounds.

\section* {Acknowledgements}

We would like to thank M. Berkooz, A. Degeratu, M. Gra\~na, J. Evslin,  I. Melnikov, N. Nekrasov, S. Theisen for useful discussions.  R.M. would like to thank Max Planck Institute for gravitational physics at Potsdam  for hospitality and A. von Humboldt  foundation for support. This work is supported in part by RTN contracts  MRTN-CT-2004-005104 and  MRTN-CT-2004-512194 and by ANR grants BLAN05-0079-01 (DA and MP) and BLAN06-3-137168 (RM).

\appendix

\section{Generalized Complex Geometry}\label{sec:gen-geom}

In this appendix, in order for the paper to be self-contained and to fix notations, we will briefly recall the basic ideas
of Generalized Complex Geometry. Generalized Complex Geometry \cite{H, G} treats
 vectors and one-forms on the same footing. Given a
$d$-dimensional manifold $M$ one defines the generalized tangent
bundle $E$ as an extension of $T$ by $T^*$
\begin{equation}
   0 \longrightarrow T^*M \longrightarrow E \longrightarrow TM \longrightarrow 0 ,
\end{equation}
whose sections are the generalized vectors. Locally they are given by the sum of
a vector and a one-form
\beq
X= v+\xi = \begin{pmatrix} v \\ \xi
\end{pmatrix} \, ,
\eeq
where $v\in TM$ and $\xi\in T^*M$. They glue on the overlap of two coordinate patches 
$U_\alpha$ and $U_\beta$ as 
\beq
\label{eq:S-patchA}
   v_{(\alpha)} + \xi_{(\alpha)}
      = a_{(\alpha\beta)} v_{(\beta)}
         + \left[ a_{(\alpha\beta)}^{-T}\xi_{(\beta)}
            - i_{a_{(\alpha\beta)}v_{(\beta)}}\omega_{(\alpha\beta)}
         \right] \, ,
\eeq
where $a_{(\alpha\beta)}$ is an element of $\GL(d,\bbR)$ and 
$\omega_{(\alpha\beta)}$ is a two-form such that
$\omega_{(\alpha\beta)}=-\dd\Lambda_{(\alpha\beta)}$, where
$\Lambda_{(\alpha\beta)}$ satisfies
\begin{equation}
\label{cocycle}
   \Lambda_{(\alpha\beta)} + \Lambda_{(\beta\gamma)}
         + \Lambda_{(\gamma\alpha)}
      = g_{(\alpha\beta\gamma)}\dd g_{(\alpha\beta\gamma)}
\end{equation}
on $U_\alpha\cap U_\beta\cap U_\gamma$ and $g_{\alpha\beta\gamma}$ is a $U(1)$
element. A one-form with these properties defines a `connective structure'' of a gerbe. 
Note that shift by $\omega$ of the one-form is what make $T^*M$ non trivially fibered
over $TM$.

There is a natural metric on $E$, given by the natural pairing of vector and
one-forms. Locally it is given by
\begin{equation}
   \eta(X,X) = i_v\xi \, ,
\end{equation}
or in matrix notation
\beq
\eta= \begin{pmatrix} 0 & 1 \\ 1 & 0 \end{pmatrix} \, .
\eeq
From the above expression it is easy to see that the stabilizer of $\eta$ is $O(d,d)$,
which acts in the fundamental representation on the generalized vectors
\beq
X'=  O X = \begin{pmatrix} a & b \\ c & d \end{pmatrix}
         \begin{pmatrix} x \\ \xi \end{pmatrix} .
\eeq
However, as discussed in Section \ref{odd}, the structure group of $E$ is reduced to the subgroup of $O(d,d)$ given
by the semi-direct product $\Ggeom=G_B\rtimes\GL(d)$.

\subsection{Generalized metrics and generalized vielbeins}\label{sec:vielbeins}

In Generalized Geometry the metric $g$ and the $B$-field
combine into a single object, the generalized metric
\beq
{\cal H} = \begin{pmatrix}
         g - B g^{-1} B & B g^{-1} \\
         - g^{-1} B & g^{-1}
      \end{pmatrix} .
\eeq
One way to justify this definition is to introduce a split of the bundle
$E$ into two orthogonal $d$-dimensional sub-bundles $E=C_+\oplus
C_-$ such that the metric $\eta$ decomposes into a positive-definite
metric on $C_+$ and a negative-definite metric on $C_-$. The two sub-bundles
are defined as
\beq
C_{\pm} = \{ X \in TM \oplus T^*M  \, \,  : \, \,  X_{\pm} = v+ (B \pm g) v \} \, ,
\eeq
and have a natural interpretation in string theory compactified in a six-dimensional manifold as the right and left mover
sectors. Then the generalized metric is defined by
\beq
   {\cal H} = \left.\eta\right|_{C_+} - \left.\eta\right|_{C_-} .
\eeq

The gluing conditions on the double overlaps  for the metric and $B$-field are 
\begin{equation} \label{Bpatch}
   g_{(\alpha)} = g_{(\beta)} , \qquad
   B_{(\alpha)} = B_{(\beta)} - \dd\Lambda_{(\alpha\beta)} \, .
\end{equation}

We can also introduce generalized vielbeins. They parametrise the coset
$O(d,d)/O(d) \times O(d)$, where the local $O(d)\times O(d)$ transformations
play the same role as the local Lorentz
symmetry for ordinary vielbeins.
There are many different conventions one could use to define the generalized
vielbeins, which are connected by local transformations. In this paper
we define the generalized vielbeins by the requirement that
the metric $\eta$ and the generalized metric
$\mathcal{H}$ can be written as 
\beq
   \eta = {\cal E}^T
      \begin{pmatrix} 0  & \mathbb{I} \\  \mathbb{I} & 0  \end{pmatrix} {\cal E} ,
      \qquad
   {\cal H} = {\cal E} ^T
      \begin{pmatrix} \mathbb{I} & 0 \\ 0 &  \mathbb{I}
\end{pmatrix} {\cal E} .
\eeq
In this basis the generalized vielbein take the form
\beq
\label{eq:genvielbeinm}
{\cal E} =
\begin{pmatrix} e & 0 \\
         - \hat{e}^T B &  \hat{e}^T
\end{pmatrix} \, ,
\eeq
which is invariant under the $\Ggeom$ subgroup of $O(d,d)$ transformations.

Note that one can a priori choose a different set of
vielbeins for the left and right mover
sectors, or equivalently for $C_\pm$
\beq
\begin{aligned}
   g &= e_\pm^T e_\pm & \text{or} &&
   g_{mn} &= e^a_{\pm\,m} e^b_{\pm\,n} \delta_{ab} \, ,\\
   g^{-1} &= \hat{e}_\pm \hat{e}_\pm^T & \text{or} &&
   g^{mn} &= \hat{e}^m_{\pm\,a} \hat{e}^n_{\pm\,b} \delta^{ab} \, ,
\end{aligned}
\eeq
and $e_\pm\hat{e}_\pm=\hat{e}_\pm e_\pm=\mathbb{I}$.
Each of the two sets is acted
upon by one of the local $O(d)$ groups.
The expression for the generalized vielbein then becomes
\beq
\label{eq:sE}
   \mathcal{E} = \frac{1}{2}\begin{pmatrix}
             (e_++e_-) + (\hat{e}_+^T-\hat{e}_-^T)B &
             (\hat{e}_+^T - \hat{e}_-^T) \\
             (e_+-e_-) - (\hat{e}_+^T+\hat{e}_-^T)B &
             (\hat{e}_+^T + \hat{e}_-^T)
          \end{pmatrix} \, .
\eeq
Since the supergravity spinors transform under one or the other of the $O(d)$ groups,
it is natural to use the local $O(d) \times O(d)$ to
set $e_+=e_-$ so that the same
spin-connections appear, for instance, in the derivatives of the two
gravitini.
Explicitly, the $O(d)\times O(d)$ action has the form
\beq
\label{eq:OdOdm}
   \cE \mapsto K \cE \ , \qquad
   K = \frac{1}{2}\begin{pmatrix} O_++O_- & O_+-O_- \\
      O_+-O_- & O_++O_- \end{pmatrix} \, ,
\eeq
where $O_\pm$ are the $O(d)$ transformation acting on the vielbeins $e_\pm$.
With this choice the generalized vielbeins reduce to those
in \eqref{eq:genvielbeinm}%


\subsection{$O(d,d)$ spinors}\label{sec:spinors}

Given the metric $\eta$, the Clifford algebra on $E$ is Cliff$(d,d)$
\beq
\{\Gamma^m,\Gamma^n\}=\{\Gamma_m,\Gamma_n\}=0\ ,\quad \{\Gamma^m,\Gamma_n\}=\delta^m_n \, .
\eeq
with $m,\ n=1\dots d$. The $\Spin(d,d)$ spinors are  Majorana--Weyl. The positive and negative chirality spin
bundles, $S^\pm(E)$, are isomorphic to even and odd forms on $E$
\begin{equation}
   \Psi_\pm \in L \otimes
      \left.\Lambda^{\text{even/odd}}T^*M\right|_{U_\alpha} \, .
\end{equation}
The isomorphism is determined by the trivial line bundle $L$, whose sections are given in
terms of the  10-dimensional dilaton, $\ee^{-\phi}\in L$. $L$ is  needed in order for
the spinors to transform correctly under diffeomorphisms and $GL(d)$.
It is easy to see that, locally, the Clifford action of $X\in E$ on
the spinors can indeed be realized as an action on forms
\begin{equation}
\label{cliff}
   X\cdot \Psi
      := (v^m \Gamma_m + \xi_m \Gamma^m)\Psi
      = i_v \Psi + \xi\wedge \Psi \, ,
\end{equation}

%
%
Also, in going from one patch to another, the
patching of $E$ implies that
\begin{equation}
\label{spinorpatch}
   \Psi^\pm_{(\alpha)}
      = \ee^{\dd\Lambda_{(\alpha\beta)}}\Psi^\pm_{(\beta)} \, ,
\end{equation}
where the exponentiated action is done by wedge product. \\

An $O(d,d)$ spinor is said to be pure if it is annihilated by half of the gamma matrices
(or equivalently if its annihilator is a maximally isotropic subspace of $E$).
Any pure spinor can be represented as a wedge product of an exponentiated complex two-form
with a complex $k$-form. The degree $k$ is called type of the pure spinor, and,
when the latter is closed, it serves as a convenient way of characterising the geometry. A  pure spinor defines an $\SU(d,d)$ structure on $E$. A further reduction of the structure group to $SU(d) \times SU(d)$ is given by the existence of a pair of compatible pure spinors. Two pure spinors are said to be compatible when they have $d/2$ common annihilators.


\section{Transforming the gauge bundle in heterotic compactifications}\label{Fhet}


As discussed at the end of Section \ref{het}, in order to map the gauge fields $\fff$ of 
the two heterotic solutions considered, we should extend our transformation on the generalized 
tangent bundle (and generalized vielbeins) to the gauge bundle. T-duality and $O(n,n)$ transformations
in heterotic string have been extended to the gauge bundle by considering $O(n+16,n+16)$ transformations.
These were introduced in \cite{GRV, EGRS}.  We will follow the same 
procedure and extend our $O(d,d)$ transformation to $O(d+16,d+16)$. Basically, we have to extend every matrix considered so far by $16$ complex components to get them on a dimension $d+16$ bundle. So we define these extended quantities:
\beq
e=\left (\begin{array}{cc} e_s & 0 \\ e_g \aaa & e_g \end{array}\right ), \quad 
g= e^T e=\left (\begin{array}{cc} g_s+ \aaa^T g_g \aaa & \aaa^T g_g \\ g_g \aaa & g_g \end{array}\right ), 
\quad B=\left (\begin{array}{cc} B_s & - \aaa^T g_g \\ g_g \aaa & B_g \end{array}\right ) \, ,
\label{gene}
\eeq
where the $s$ index denotes the space-time objects (they are the same as in Section \ref{subgenviel}),
and the $g$ index denotes the gauge bundle quantities. 
$\aaa$ is the $16\times d$ matrix giving the gauge connection. $g_g=e_g^T e_g$ and $B_g$ are the ``gauge'' 
metric and $B$-field, which are actually constrained to take specific values, 
in order to make sense with the (root) lattice on which we consider the fields
\beq
g_g=\frac{1}{2} \mathcal{C}, \quad (B_{g})_{ij}=\left \{\begin{array}{cc} -(g_g)_{ij} & i<j \\ 0 & i=j \\ (g_g)_{ij} & i>j \end{array}\right.
\eeq
where $\mathcal{C}$ is the Cartan matrix (symmetric) of the group considered. As these matrices are fixed, 
the only new freedom we introduce is the gauge connection given by $\aaa$.

Then we define as before the generalized metric $\mathcal{H}$ and the generalized vielbein $\mathcal{E}$, 
which are now extended to the gauge bundle:
\beq
\tilde{\mathcal{E}}=\left (\begin{array}{cc} e & 0 \\ -\hat{e} \, B & \hat{e} \end{array}\right ) \qquad  
\mathcal{H}=\tilde{\mathcal{E}}^T\ \tilde{\mathcal{E}}=\left (\begin{array}{cc} g-Bg^{-1}B & Bg^{-1} \\ -g^{-1}B & g^{-1} 
\end{array}\right ) \ , 
\eeq
and are therefore $2(d+16)\times 2(d+16)$ matrices. The $O(d+16,d+16)$ transformations act on them as did 
$O(d,d)$ on the generalized vielbein and metric, \eqref{eq:genvfiber0} and \eqref{eq:genmetric}. 
We define the transformation of the dilaton as before (\ref{dil0}); as we will see, we can use either the previous $d\times d$ metric or the new $(d+16)\times (d+16)$ one. 

As in \cite{GRV,EGRS}, we shall consider a subset of $O(d+16,d+16)$, which does not 
change $e_g$ and $B_g$. 
Indeed, $e_g$ and $B_g$ are related to the Cartan matrix which should stay invariant. 
Furthermore, the transformation should preserve the off-diagonal structure of $B$, i.e. 
the off-diagonal block of the transformed $B$ should be related in the same way to the 
new gauge connection. \\

Following the logic of Section \ref{odd}, we consider the following $O(d+16,d+16)$ transformations
\beq
\label{Od16tra}
O=\left (\begin{array}{cc} A & 0 \\ C & A^{-T} \end{array}\right ), \\
\eeq
which satisfies the $O(d+16,d+16)$  constraint $A^T C+C^T A=0_{d+16}$, and 
where, according to \eqref{gene}, the matrices $A$ and $C$ can be decomposed into geometric and gauge blocks
\beq
A= \begin{pmatrix} A_s & 0 \\ A_o & A_g \end{pmatrix}, \ C= \begin{pmatrix} C_s & C_o \\ C^\prime_o & C_g
\end{pmatrix} \, .
\eeq
The transformed vielbeins read
\beq
\tilde{\mathcal{E}}^\prime= \begin{pmatrix} e^\prime & 0 \\ -\hat{e}^{\prime} \, B^\prime & \hat{e}^{\prime} 
\end{pmatrix}, \quad e^\prime=e A, \quad B^\prime=A^T B A - A^T C \ .
\label{tran}
\eeq

Imposing the invariance of the $e_g$ component of the vielbeins sets $A_g= \mathbb{I}_{16}$ and
gives the new gauge connection $\mathcal{A}^\prime = \mathcal{A} A_s + A_o$. Similarly the invariance
of $B_g$ in the  $B$-field implies $C_g=0_{16}$. Then we have to ask that the off-diagonal
terms in $B$ can be written again in the form \eqref{gene}. This fixes $C_o$ and 
\beq
C_o = A_s^{-T}A_o^T(g_g+B_g) \qquad  C^\prime_o = (B_g-g_g)A_o 
\eeq

Finally it is easy to see that the $O(d+16,d+16)$  constraint $A^T C+C^T A=0_{d+16}$ is 
equivalent to the antisymmetry of transformed $B$-field in \eqref{tran} and gives the constraint
\beq
\label{constr}
A_s^T \, C_s + C_s^T \, A_s=2 A_o^T g_g A_o \ .
\eeq


\subsection{A specific case: the K\"ahler-non K\"ahler transition of Section \ref{het}}

Let us now focus on the specific examples we considered in Section \ref{het}. Solution 1 is a trivial
$\mathbb{T}^2$ fibration, with no $B$-field, so we set $B_s=0$, and 
has a non trivial
gauge connection $\aaa \neq 0$. To recover Solution 2, we want to produce a connection in the metric,
a non-zero $B$-field, and no gauge connection, i.e. $\aaa^\prime=0$. From Section \ref{odd}, 
it is easy to write the metric part of the transformation $A$
\beq
A_s= \begin{pmatrix} \mathbb{I}_4 & 0 \\ A_{\ccc} & \mathbb{I}_2 \end{pmatrix} \, .
\eeq
Since the diagonal elements are just identity matrices, this transformation does not modify the metric 
and the dilaton. The vanishing of the gauge field $\aaa^\prime=0$ simply tells us to choose 
$A_o=- \aaa A_s$. So the choice of connections fixes completely the $A$ matrix. 

We have now to check whether the constraint (\ref{constr}) can be satisfied. 
If we take the gauge connection in Solution 1 to be only on the base, the
off-diagonal block in the vielbein \eqref{gene}  takes the form 
$e_g \aaa= \begin{pmatrix} \aaa_{\mathcal{B}} & 0_{16\times 2} \end{pmatrix}$, then 
the constraint \eqref{constr} becomes
\beq
A_s^TC_s+C_s^TA_s= 2 \aaa^T g_g \aaa \ ,
\eeq
and it is easy to verify that it solved by the following choice for the matrix $C_s$
\beq
C_s=\left (\begin{array}{cc} \tilde{C}_{\bbb}-A_{\ccc}^T C_{\ccc}+\aaa_{\mathcal{B}}^T \aaa_{\mathcal{B}} & -(C_{\ccc}^T+A_{\ccc}^T C_{\fff}) \\ C_{\ccc} & C_{\fff} \end{array}\right ) \ ,
\eeq
where $\tilde{C}_{\bbb}$, $C_{\fff}$ and $C_{\ccc}$ are free, and the two first are antisymmetric. Note the new $B$-field is then given by
\beq
B_s^\prime=-\left (\begin{array}{cc} \tilde{C}_{\bbb} & -C_{\ccc}^T \\ C_{\ccc} & C_{\mathcal{F}} \end{array}\right ) \label{bp}
\eeq
so we see once again that we can choose it to be whatever we want, and it fixes completely the $C$ matrix.\\

To summarise, inspired by the T-duality in heterotic strings we have made some steps towards extending the $O(d,d)$ generalized tangent bundle transformations to $O(d+16,d+16)$ hence covering the transformations of the gauge bundle. This allows, in particular, to relate the two solutions discussed in Section \ref{het}.


\begin{thebibliography}{45}

\bibitem{GPR} A. Giveon, M. Porrati, E. Rabinovici, \textit{Target Space Duality in String Theory}, \textit{Phys. Rept.} {\bf 244} (1994) 77 [hep-th/9401139]

\bibitem{GP} M. Gra\~na, J. Polchinski, \textit{Gauge / gravity duals with holomorphic dilaton}, \textit{Phys. Rev. D} {\bf 65} (2002) 126005 [hep-th/0106014]

\bibitem{KSTT} S. Kachru, M. B. Schulz, P. K. Tripathy, S. P. Trivedi, \textit{New supersymmetric string compactifications}, \textit{JHEP} {\bf 0303} (2003) 061 [hep-th/0211182]

\bibitem{Sc} M. B. Schulz, \textit{Superstring orientifolds with torsion: O5 orientifolds of torus fibrations and their massless spectra}, \textit{Fortsch. Phys.} {\bf 52} (2004) 963 [hep-th/0406001]

\bibitem{GMPT06} M. Gra\~na, R. Minasian, M. Petrini, A. Tomasiello, \textit{A scan for new $\mathcal{N}=1$ vacua on twisted tori}, 
\textit{JHEP} \textbf{0705} (2007) 031 [hep-th/0609124]

\bibitem{S} A. Strominger, \textit{Superstrings with torsion}, \textit{Nucl. Phys. B} \textbf{274} (1986) 253

\bibitem{Hu} C. M. Hull, \textit{Superstring Compactifications With Torsion And Space-Time Supersymmetry}, Print-86-0251 (CAMBRIDGE)

\bibitem{DRS} K. Dasgupta, G. Rajesh, S. Sethi, \textit{M Theory, Orientifolds and $G$-flux}, \textit{JHEP} {\bf 9908} (1999) 023 [hep-th/9908088]

\bibitem{BD} K. Becker, K. Dasgupta, \textit{Heterotic Strings with Torsion}, \textit{JHEP} {\bf 0211} (2002) 006 [hep-th/0209077]

\bibitem{GoldP} E. Goldstein, S. Prokushkin, \textit{Geometric model for complex non-Kaehler manifolds with SU(3) structure}, \textit{Commun. Math. Phys.} {\bf 251} (2004) 65 [hep-th/0212307]

\bibitem{FuY} J-X. Fu, S-T. Yau, \textit{The Theory of superstring with flux on non-Kahler manifolds and the
  complex Monge-Ampere equation}, [hep-th/0604063]
  
\bibitem{BBFTY} K. Becker, M. Becker, J-X. Fu, L-S. Tseng, S-T. Yau, \textit{Anomaly Cancellation and Smooth Non-K\"ahler Solutions in Heterotic String Theory}, \textit{Nucl. Phys. B} {\bf 751} (2006) 108 [hep-th/0604137]

\bibitem{Yi1} T. Kimura, P. Yi, \textit{Comments on heterotic flux compactifications}, \textit{JHEP} {\bf 0607} (2006) 030 [hep-th/0605247]

\bibitem{Yi2} S. Kim, P. Yi, \textit{A heterotic flux background and calibrated five-branes}, \textit{JHEP} {\bf 0611} (2006) 040 [hep-th/0607091]

\bibitem{BTY} M. Becker, L-S. Tseng, S-T. Yau, \textit{Heterotic K\"ahler/non-K\"ahler Transitions}, [arXiv:0706.4290]

\bibitem{Se} S. Sethi, \textit{A Note on Heterotic Dualities via M-theory}, \textit{Phys. Lett. B} {\bf 659} (2008) 385 [arXiv:0707.0295]

\bibitem{H} N. Hitchin, \textit{Generalized Calabi-Yau manifolds}, [math/0209099]

\bibitem{G} M. Gualtieri, \textit{Generalized complex geometry}, Oxford University DPhil thesis, [math.DG/0401221]

\bibitem{GMPT04} M. Gra\~na, R. Minasian, M. Petrini, A. Tomasiello, \textit{Supersymmetric backgrounds from generalized Calabi-Yau manifolds}, \textit{JHEP} \textbf{0408} (2004) 046 [hep-th/0406137]

\bibitem{GMPT05} M. Gra\~na, R. Minasian, M. Petrini, A. Tomasiello, \textit{Generalized structures of $\mathcal{N}=1$ vacua}, \textit{JHEP} \textbf{0511} (2005) 020 [hep-th/0505212]

\bibitem{H2} N. Hitchin, \textit{Brackets, forms and invariant functionals}, [math/0508618]

\bibitem{Hu2} C. M. Hull, \textit{Generalized Geometry for M-theory}, \textit{JHEP} {\bf 0707} (2007) 079 [hep-th/0701203]

\bibitem{PW} P. P. Pacheco, D. Waldram, \textit{M-theory, exceptional generalized geometry and superpotentials}, \textit{JHEP} {\bf 0809} (2008) 123 [arXiv:0804.1362]

\bibitem{GLSW} M. Gra\~na, J. Louis, A. Sim, D. Waldram, \textit{E7(7) formulation of N=2 backgrounds}, [arXiv:0904.2333]

\bibitem{LM} O. Lunin, J. M. Maldacena, \textit{Deforming field theories with $U(1) \times U(1)$ global symmetry and their gravity duals}, \textit{JHEP} {\bf 0505} (2005) 033 [hep-th/0502086]

\bibitem{MPZ} R. Minasian, M. Petrini, A. Zaffaroni, \textit{Gravity duals to deformed SYM theories and Generalized Complex Geometry}, \textit{JHEP} \textbf{0612} (2006) 055 [hep-th/0606257]

\bibitem{HT} N. Halmagyi, A. Tomasiello, \textit{Generalized K\"ahler Potentials from Supergravity}, [arXiv:0708.1032]

\bibitem{LT} D. Lust, D. Tsimpis, \textit{Supersymmetric AdS(4) compactifications of IIA supergravity}, \textit{JHEP} \textbf{0502} (2005) 027 [hep-th/0412250]

\bibitem{KT} P. Koerber, D. Tsimpis, \textit{Supersymmetric sources, integrability and generalized-structure compactifications}, \textit{JHEP} \textbf{0708} (2007) 082 [arXiv:0706.1244]

\bibitem{GMPW} M. Gra\~na, R. Minasian, M. Petrini, D. Waldram, \textit{T-duality, Generalized Geometry and Non-Geometric Backgrounds}, [arXiv:0807.4527]

\bibitem{An} D. Andriot, \textit{New supersymmetric flux vacua with intermediate SU(2) structure}, \textit{JHEP} \textbf{0808} (2008) 096 [arXiv:0804.1769]

\bibitem{CG} G. R. Cavalcanti, M. Gualtieri, \textit{Generalized complex structures on nilmanifolds}, [math/0404451]

\bibitem{CHSW} P. Candelas, G. T. Horowitz, A. Strominger, E. Witten, \textit{Vacuum configurations for superstrings}, \textit{Nucl. Phys.} \textbf{B 258} (1985) 46

\bibitem{swann} A. Swann, \textit{Twisting Hermitian and hypercomplex geometries}, [arXiv:0812.2780]

\bibitem{A} P.S. Aspinwall, \textit{An analysis of fluxes by duality}, [hep-th/0504036]

\bibitem{EM} J. Evslin, R. Minasian, \textit{Topology Change from (Heterotic) Narain T-Duality}, [arXiv:0811.3866]

\bibitem{Al} A. Alekseev, T. Strobl, \textit{Current algebras and differential geometry}, \textit{JHEP} {\bf 0503} (2005) 035 [hep-th/0410183]

\bibitem{BHM} D. M. Belov, C. M. Hull, R. Minasian, \textit{T-duality, Gerbes and Loop Spaces}, [arXiv:0710.5151]

\bibitem{GRV} A. Giveon, E. Rabinovici, G. Veneziano, \textit{Duality in String Background Space}, \textit{Nucl. Phys. B} \textbf{322} (1989) 167\\
A. Giveon, M. Ro\v{c}ek, \textit{Generalized duality in curved string backgrounds}, \textit{Nucl. Phys. B} \textbf{380} (1992) 128 [hep-th/9112070]

\bibitem{EGRS} S. Elitzur, E. Gross, E. Rabinovici, N. Seiberg, \textit{Aspects of bosonization in string theory}, \textit{Nucl. Phys. B} \textbf{283} (1987) 413

\end{thebibliography}
\end{document}